\documentclass[a4paper,11pt]{article}
\pdfoutput=1 

\usepackage{jheppub} 

\usepackage[T1]{fontenc} 

\title{\boldmath Influence of quintessence dark energy on the shadow of black hole}

\author[a,b]{Xiao-Xiong Zeng,}
\author[c,d]{Hai-Qing Zhang}

\affiliation[a]{State Key Laboratory of Mountain Bridge and Tunnel Engineering, Chongqing\\ Jiaotong University, Chongqing 400074, China}
\affiliation[b]{Department of Mechanics, Chongqing Jiaotong University, Chongqing ~400074, China}
\affiliation[c]{Center for Gravitational Physics, Department of Space Science, Beihang University, Beijing 100191, China.}
\affiliation[d]{International Research Institute for Multidisciplinary Science, Beihang University, Beijing 100191, China}

\emailAdd{xxzengphysics@163.com}
\emailAdd{hqzhang@buaa.edu.cn}

\abstract{We investigate the effects of quintessence dark energy on the shadows of black hole, surrounded by various profiles of accretions.  For the thin disk accretion, the images of the black hole comprises the dark region and bright region, including direct emission, lensing rings and photon rings. Although their details depend on the form of the emission, generically, the direct emission plays a major role for the observed brightness of the black hole, while the lensing ring makes a small contribution and photon ring makes a negligible contribution. The existence of cosmological horizon also plays an important role in the shadows, since the observer in the domain of outer communications is nearby the cosmological horizon. For the spherically symmetric accretion, the static and infalling matters are considered. We find that the positions of photon spheres are the same for both static and infalling accretions. However, the observed specific intensity of the image for the infalling accretion is darker than the static accretion, due to the Doppler effect of the infalling movement.
}

\begin{document}
\maketitle
\flushbottom

\section{Introduction}
\label{sec:intro}
The Event Horizon Telescope (EHT) collaboration has recently released an image of a supermassive black hole in M87 \cite{Akiyama:2019cqa,Akiyama:2019brx,Akiyama:2019sww,Akiyama:2019bqs,Akiyama:2019fyp,Akiyama:2019eap}. The image shows that there is a dark interior surrounding by a bright ring, which are called the black hole shadow and photon ring, respectively. The shadow and photon ring are resulted from the deflections of light, or gravitational lensing by the black holes, a prediction from Einstein's General Relativity (GR) \cite{Wald:1984rg}. Therefore, the shadow image of black hole in M87 from EHT is another strong evidence for GR.

The deflection of light by an intense star or black holes was initiated in \cite{Synge:1966okc, Bardeen:1972fi}. Later, it was extended to a black hole surrounded by a thin accretion disk \cite{Luminet:1979nyg}. According to these papers, there exists a critical curve in the observed image of the black hole. When traced backwards from the distant observer, the light ray from this critical curve will asymptotically approach a bound photon orbit. This bound orbits for Schwarzschild black hole is $r=3M$ ($M$ is the mass of the black hole), and the critical curve has the radius of $b=\sqrt{27}M$ ($b$ is called impact parameter).  Thus, the shadow of the black hole represents the dark interior of the critical curve. However, according to \cite{Gralla:2019xty}, the radius and the profile of the shadow would depend on the location and form of the accretion surrounding the black hole.  Besides the shadow, there also exists the enhanced brightness, called photon ring, around the black hole. This is because the light rays near the critical curve may orbit many times of the black hole, thus picking up extra brightness.  So far, the study of shadows of black holes has become a subject of great interest \cite{Jaroszynski:1997bw,Falcke:1999pj, Bambi:2013nla, Shaikh:2018lcc, Narayan:2019imo, Gralla:2019drh, Allahyari:2019jqz, Li:2020drn, Banerjee:2019nnj, Vagnozzi:2019apd, Vagnozzi:2020quf, Safarzadeh:2019imq, Davoudiasl:2019nlo, Roy:2019esk, Chen:2019fsq, Cunha:2019hzj,  Konoplya:2020bxa, Roy:2020dyy, Islam:2020xmy, Jin:2020emq, Guo:2020zmf, Wei:2020ght, Zeng:2020dco, Caiyifu,Khodadi:2020jij, Cuadros-Melgar:2020kqn, Konoplya:2019sns,Zhang:2019glo}. In particular,  the shadows of a high-redshift supermassive black hole may serve as the standard rulers \cite{Vagnozzi:2020quf}, which would constrain the cosmological parameters. Moreover, the EHT data may impose constraints in particle physics via the mechanism of superradiance   \cite{Davoudiasl:2019nlo, Roy:2019esk}. More references can be referred to \cite{Shaikh:2018lcc}.

Recent astronomical observations show that the Universe is accelerating expansion \cite{Perlmutter, Riess, Garnavich}, implying a state of negative pressure. Quintessence dark energy is one of the candidates to interpret the negative pressure. The state equation of quintessence dark energy is governed by the pressure $p$ and the energy density $\rho$ with $p=w\rho$, where $w$ is the state parameter. It is deemed that the range $-1<w<-1/3$ causes the acceleration of the Universe. Black holes in quintessence dark energy models were extensively studied \cite{Kiselev:2002dx}. To investigate the shadows of black hole in the quintessence dark energy model will bring new insights and impose restrictions to the quintessence model. Previous work on shadows in dark energy model can be found in \cite{Lacroix:2012nz,Khan:2020ngg}.

In this paper, we will investigate the shadow images of quintessence black holes with different accretions. In particular, we study the geometrically thin and optically thin disk accretion and spherically symmetric accretion. Apparently different accretions will lead to different shadow images. Specifically, for the thin disk accretion, the images of the black hole comprises the shadows, lensing rings and photon rings. The laters are resulted from the intersections of the light rays and the thin accretion for many times.  Although different emission profiles of the accretion is important to the observed specific intensities, the direct image from accretion makes major contributions to the brightness of the image. Besides, the lensing ring makes small contributions and the photon ring makes negligible contributions to the observed specific intensities. For the spherically symmetric accretions, the shadows and photon rings are also spherically symmetric. The static accretion and an infalling accretion will have different impacts to the observed specific intensities, although the radii of the shadows are not changed. Specifically, the infalling accretion will have extra Doppler effect, which will lead to darker shadows than the static case.  A key point in the cosmological black hole is that there exists a domain of outer communication  between the event horizon and cosmological horizon \cite{Friedman, Galloway}. Therefore, various quintessence state parameter $w$ will lead to different distances between the event horizon and cosmological horizon. This will certainly have impact on the distance between the observer (near cosmological horizon) and the accretion (near the event horizon). Therefore, different quintessence state parameters will lead to different forms of the observed specific intensities.  We expect that the shadow images of the quintessence black hole will impose constraints of the quintessence dark energy model in the Universe from the observation of EHT.

This paper is arranged as follows: In section \ref{light} the black hole solutions surrounded by quintessence matter and the light deflections by the quintessence black hole are  introduced; In section \ref{disk} the shadow images of the black hole with thin disk accretion are given;   In section \ref{spherical}, we show the images of the black hole with spherically symmetric accretions;  Finally, we draw the conclusions and discussions in section \ref{conclusion}.

\section{Light deflection by a  black hole in quintessence dark energy}
\label{light}

The geometry of  a black hole in a quintessence dark energy model can be expressed as
\begin{equation}
ds^2 = -f(r) dt^2 + f(r)^{-1} dr^2 +  r^2 ( d \theta^2 + sin^2\theta d \psi^2), \label{metric}
\end{equation}
where,
\begin{equation}
f(r)= 1 - \frac{2 M} {r}- \frac{a}{ r^{ 3 w + 1} },  \label{ffr}
\end{equation}
in which, $M$ is the mass of the black hole,  $a$ and $w$ are the normalization factor and state parameter respectively. For the case of $w= -1$, the metric in Eq.(\ref{ffr}) reduces  to the Schwarzschild-de-Sitter black hole with
\begin{equation}
f(r)= 1 - \frac{ 2 M} { r}- a r^2,
\end{equation}
which has both the event horizon  $r_{h}$  and cosmological horizon $r_{c}$. For $w = -\frac{2}{3}$, we can obtain the event  horizon  and cosmological horizon as
\begin{equation}
r_{h} = \frac{ 1 - \sqrt{ 1 - 8 M a} }{ 2 a},
~~~
r_{c} = \frac{ 1 + \sqrt{ 1 - 8 M a} }{ 2 a},
\end{equation}
where the  restriction $ 8 M a <1$ has been imposed.  For $w= -1/3$, the   metric in Eq.(\ref{ffr}) reduces  to the  Schwarzschild black hole, which has only an event horizon.
In this paper, we are interested in the quintessence
dark energy model, in which the state parameter  $w$ is  restricted in the region  $-1 < w < -\frac{1}{3}$, and there always exists an event horizon and a cosmological horizon. The region between the two
horizons is called the domain of outer communication \cite{Friedman, Galloway}, since
any two observers in this region may communicate
with each other without being hindered by a horizon.

The equation of state for the quintessence  dark energy  is
\begin{equation}
    p = w \rho,~~~\rho = - \frac{a} {2}\frac{3 w}{r^{3(1+w)}}
\end{equation}
in which  $p$ is the pressure while $\rho$ is the energy density.  In order to cause cosmic acceleration, the pressure of the quintessence  dark energy  $p$  should  be  negative, thus the energy density $\rho$ is positive if the normalization factor  $a$ is positive.  In this paper, we will only  focus  on the effect of   state parameter on the shadows of a quintessence black hole, and the we will set the normalization factor $a=0.05$ and black hole $M=1$ thereafter.

In order to investigate the light deflection in the
 quintessence  black hole, we need to find how the light ray moves in this spacetime. We will resort to the following Euler-Lagrange equation
\begin{equation}
\frac{d}{ds}\left(\frac{\partial \mathcal{L}}{\partial \dot{x}^{\mu}}\right)=\frac{\partial \mathcal{L}}{\partial x^{\mu}},
\label{eleq}
\end{equation}
in which $s$ is the affine parameter of the light trajectory, the symbol ~$\dot{}$~ is the derivative with respect to $s$, thus $\dot{x}^{\mu}$ the four-velocity of the light ray and  $\mathcal{L}$ is the Lagrangian, taking the form as
\begin{equation}
 \mathcal{L}=\frac{1}{2}g_{\mu\nu}\dot{x}^{\mu}\dot{x}^{\nu}=\frac{1}{2}\left( - f(r)\dot{t}^2+\frac{\dot{r}^2}{f(r)}+r^2\left(\dot{\theta}^2+\sin^2{\theta}~ \dot{\psi}^2 \right)\right).
\label{laeq}
 \end{equation}
We will also impose the initial conditions $\theta = \pi/2$, $\dot{\theta} = 0$. That is,  the light rays always   move in the equatorial plane.  In addition,  since the metric coefficients do not depend explicitly on time $t$ and azimuthal angle $\psi$, there are two  conserved quantities correspondingly.
  Combining Eqs.(\ref{ffr}), (\ref{eleq}) and  (\ref{laeq}), the time, azimuthal and radial components of the four-velocity satisfy the following equations of motions
\begin{eqnarray}
  &&\dot{t}=\frac{1}{b \left(1 - \frac{2 M} {r}- \frac{a}{ r^{ 3 w + 1} }\right)},  \label{time} \\
  &&\dot{\psi}=\pm \frac{1}{r^2}, \label{psi} \\
  &&\dot{r}^2+\frac{1}{r^2} \left(1 - \frac{2 M} {r}- \frac{a}{ r^{ 3 w + 1} }\right)= \frac{1}{b^2},
 \label{radial}
\end{eqnarray}
where $+$ and $-$  in Eq.\eqref{psi} correspond to the counterclockwise and clockwise direction respectively for the motion of light rays. The parameter $b={|L|}/{E} $ is the  impact parameter, in which  $E$ and $L$ are the conserved quantities with respect to the time and azimuthal direction. Moreover, Eq.(\ref{radial}) can be rewritten as
\begin{equation}
 \dot{r}^2+V(r)=\frac{1}{b^2},  \label{vbr}
\end{equation}
where
\begin{equation}
V(r)=\frac{1}{r^2} \left(1 - \frac{2 M} {r}- \frac{a}{ r^{ 3 w + 1} }\right),
\end{equation} \label{epotential}
is an effective potential.
At the  photon sphere, the motions of the light ray   satisfy  $\dot{r}=0$, and  $\ddot{r}=0$, which also means
\begin{eqnarray}
V(r)=\frac{1}{b^2}, ~~~ V^{'}(r) = 0,
\label{condition1}
\end{eqnarray}
where the  prime $'$ denotes the derivative with respect to the radial coordinate $r$. Based on Eq.(\ref{condition1}), we can obtain the radius and  impact parameter of the  photon sphere for different $w$. Generically, it is difficult to obtain the analytic results of the radius and impact parameter. But
for $w=-2/3$ the metric function can be simplified, thus the radius and  impact parameter of the  photon sphere can be expressed as
 \begin{eqnarray}
  &&r_{ph}= \frac{1-\sqrt{1-6 a M}}{a},\\
 && b_{ph}=   \frac{\sqrt{2} \sqrt{6 M \sqrt{1-6 a M}-\frac{\sqrt{1-6 a M}}{a}+\frac{1}{a}-9 M}}{\sqrt{8 a^2 M-a}}.
 \end{eqnarray}
 For other  $w$'s, the numerical results of radii and impact parameters of the photon sphere, as well as the event horizons and cosmological horizons are  listed  in the Table \ref{tab11}. From this table, we see that as one increases the absolute value of $w$,  the event horizon $r_h$ and   impact parameter $b_{ph}$ increase as well, while the cosmological horizon $r_c$ decreases. The radius of photon sphere however is non-monotonic, it increases at first and then decreases around $w=-0.7$.
In addition, from Table \ref{tab11} we see that the locations of the event horizon and cosmological horizon become closer as  $|w|$ increases, which leads to a narrower domain of the outer communication.

\begin{table}[h]
\begin{center}
\caption{The values of radius $r_{ph}$ and impact parameter $b_{ph}$ of the photon sphere, as well as the event horizon $r_h$ and cosmological horizon  $r_c$    for various $w$'s with $M=1$,  $a=0.05$.}\label{tab11}
\vspace{2mm}
\begin{tabular}{ccccccc}
\hline       &{$w=-0.4$}      & {$w=-0.5$}     & {$w=-0.6$} &   {$w=-0.7$} &  {$w=-0.8$} & {$w=-0.9$} \\    \hline
{$r_{ph}$} &  {3.18038}      &{3.21631}           &{3.25038}     &{3.27103}          &{3.25476}     &{3.16885}      \\
{$b_{ph}$}  & {5.72947}      &{5.98805}        &{6.42035}      &{7.23434}              &{9.22055}             &{27.141} \\
{$r_{h}$}  & {2.12343}     &{2.15857}         &{2.20806}       &{2.28304}              &{2.41487}             &{2.8274} \\
{$r_{c}$}  & {3199990}     &{395.969}         &{39.6448}       &{13.1027}              &{6.55098}             &{3.65555} \\ \hline
\end{tabular}
\end{center}
\end{table}

\begin{figure}[tbp]
\centering 
\includegraphics[width=.42\textwidth]{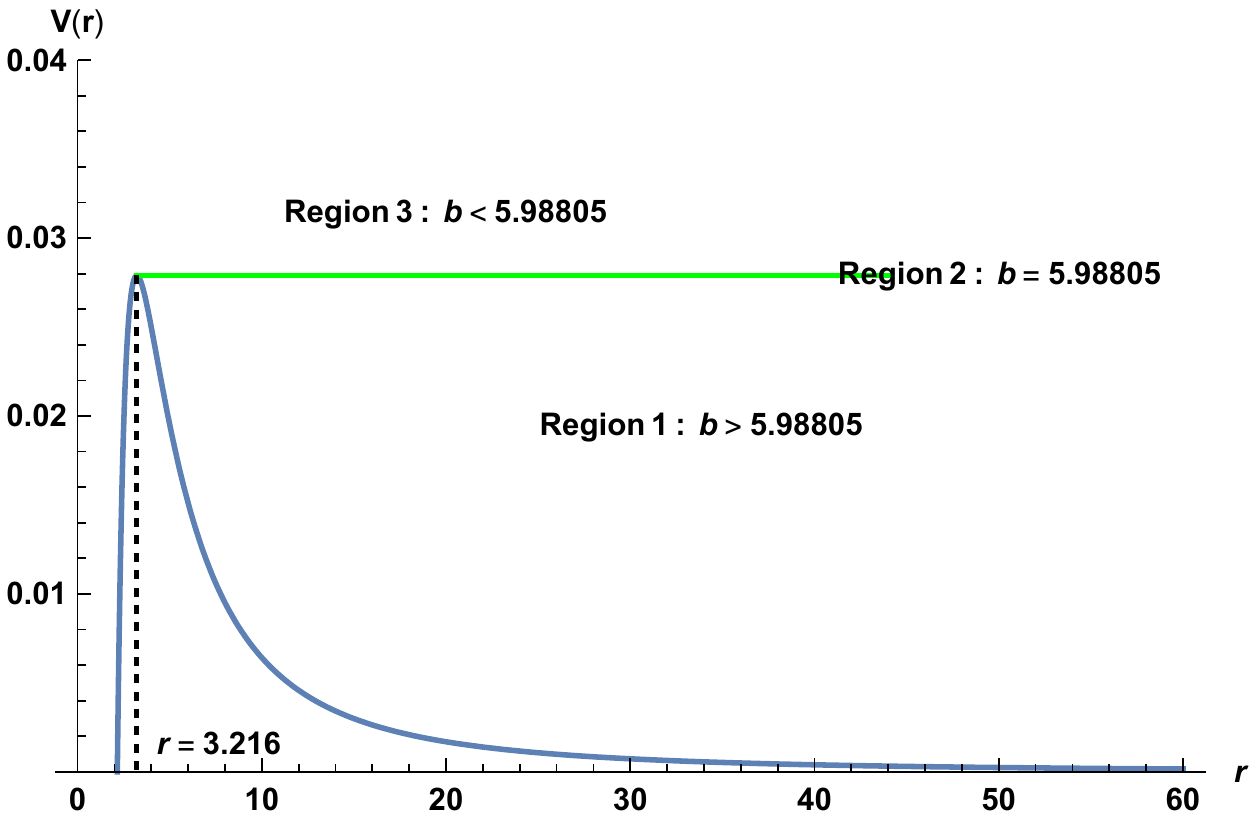}
\hfill
\includegraphics[width=.42\textwidth]{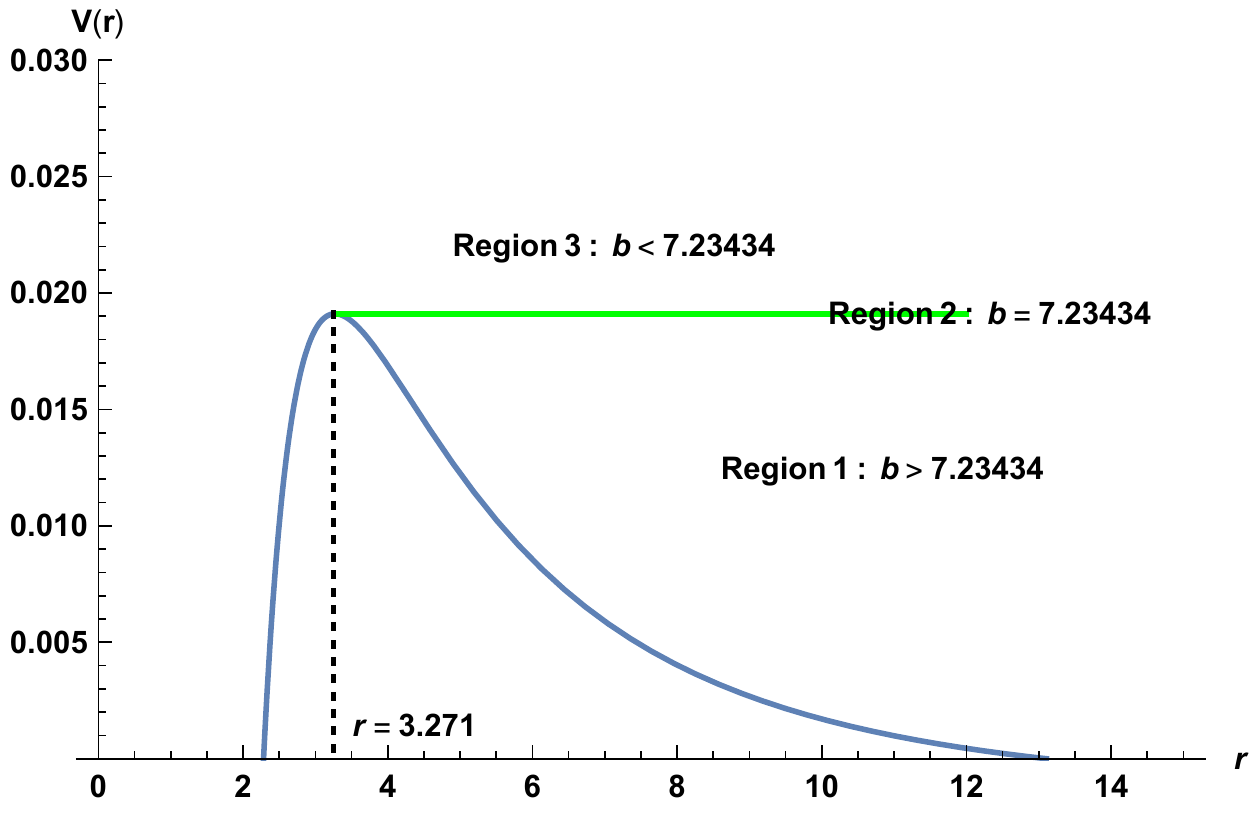}
\caption{\label{fig1}   The profile of the effective potential (blue lines) for $w=-0.5$ (left panel) and $w=-0.7$ (right panel) with $M=1$ and $a=0.05$. The dashed lines indicate the radii of the photon sphere $r_{ph}$. Region 2 (green lines) correspond to $V(r)=1/b_{ph}^2$,   while  Region  1  and  Region 3  correspond to   $V(r)<1/b_{ph}^2$ and  $V(r)>1/b_{ph}^2$ respectively. }
\end{figure}

From the Eq.(\ref{vbr}), we see that the motion of the particle depends on the impact parameter and the effective potential. In Figure \ref{fig1}, the effective potential is plotted  for different state parameters $w=-0.5$ and $w=-0.7$.  We can see that at the event horizon, the effective
potential vanishes. Then it increases and reaches a maximum at the photon sphere $r_{ph}$, and later decreases and vanishes at the cosmological horizon. Let's consider a light ray moves in the radially inward direction.
In Region 1, if the light ray starts its
motion at $r>r_{ph}$, the light ray will encounter the potential barrier and be reflected back to the outward direction. If the photon starts the motion at $r<r_{ph}$,  the photons will fall into the singularity. In Region 2, namely  $b=b_{ph}$, as the light ray approaches the photon sphere, it will revolve around the black hole infinitely many times since the angular velocity is non-zero. Actually, this orbit is  circular and unstable \cite{Gralla:2019xty}. In Region 3, the light ray will continue moving in the inward direction since it does not encounter the potential barrier. Finally it will enter the inside of the black hole and  fall into the singularity.

Since we have already obtained the effective potential,  it is also convenient to obtain the effective force on the photon, which is \footnote{Here, we have divided $dV/
dr$ by 2 since the equation of motion  has been written as    Eq.(\ref{vbr}). }
\begin{equation}
F = -\frac{1}{2} \frac{ d V(r)}{dr} =-\frac{3 M}{r^4}+\frac{1}{r^3}-\frac{1}{2} 3 a (w+1) r^{-3 w-4},
\end{equation}
The first term $-{3 M}/{r^4}$, called the Newtonian term, is attractive as it is negative while the second term $1/r^3$ is repulsive. The contributions of the quintessence dark energy comes from the third term $-\frac{1}{2} 3 a (w+1) r^{-3 w-4}$, which plays the role of an attractive force.

The trajectory of the light ray can be depicted on the basis of the equation of motion.   Combining  Eqs.(\ref{psi}) and (\ref{radial}), we have
\begin{equation}
\frac{dr}{d\psi}=\pm r^2 \sqrt{\frac{1}{b^2}-\frac{1}{r^2}\left(1 - \frac{2 M} {r}- \frac{a}{ r^{ 3 w + 1} }\right)}. \label{drp}
\end{equation}
In order to   integrate it out  conveniently,  we set   $u=1/r$. Thus, Eq.(\ref{drp})  becomes
\begin{equation}
\frac{du}{d\psi}=\sqrt{\frac{1}{b^2}-u^2 \left(-a u^{3 w+1}-2 M u+1\right)}\equiv \Phi(u).\label{gu}
\end{equation}
The geometry of the geodesics depends on the  roots of the equation $ \Phi(u) =0$. In particular, for $ b>b_{ph}$,   light will be deflected at the radial position $ u_i$ which satisfies $ \Phi(u_i) =0$. Therefore, it is important to find the  radial position $u_i$ in order to obtain the trajectory of the light ray. In addition, the location of the observer is also important. Usually, the observer is located at infinite boundary for the asymptotically flat spacetime. However, for the model of quintessence
black hole in our paper, the cosmological horizon is present. Physically, the observer should be located in the domain of outer communication which is between the event horizon and cosmological horizon, just as in de Sitter spacetime. For convenience, the observer in our paper is set to be near the cosmological horizon.


\begin{figure}[h]
\centering 
\includegraphics[width=.45\textwidth]{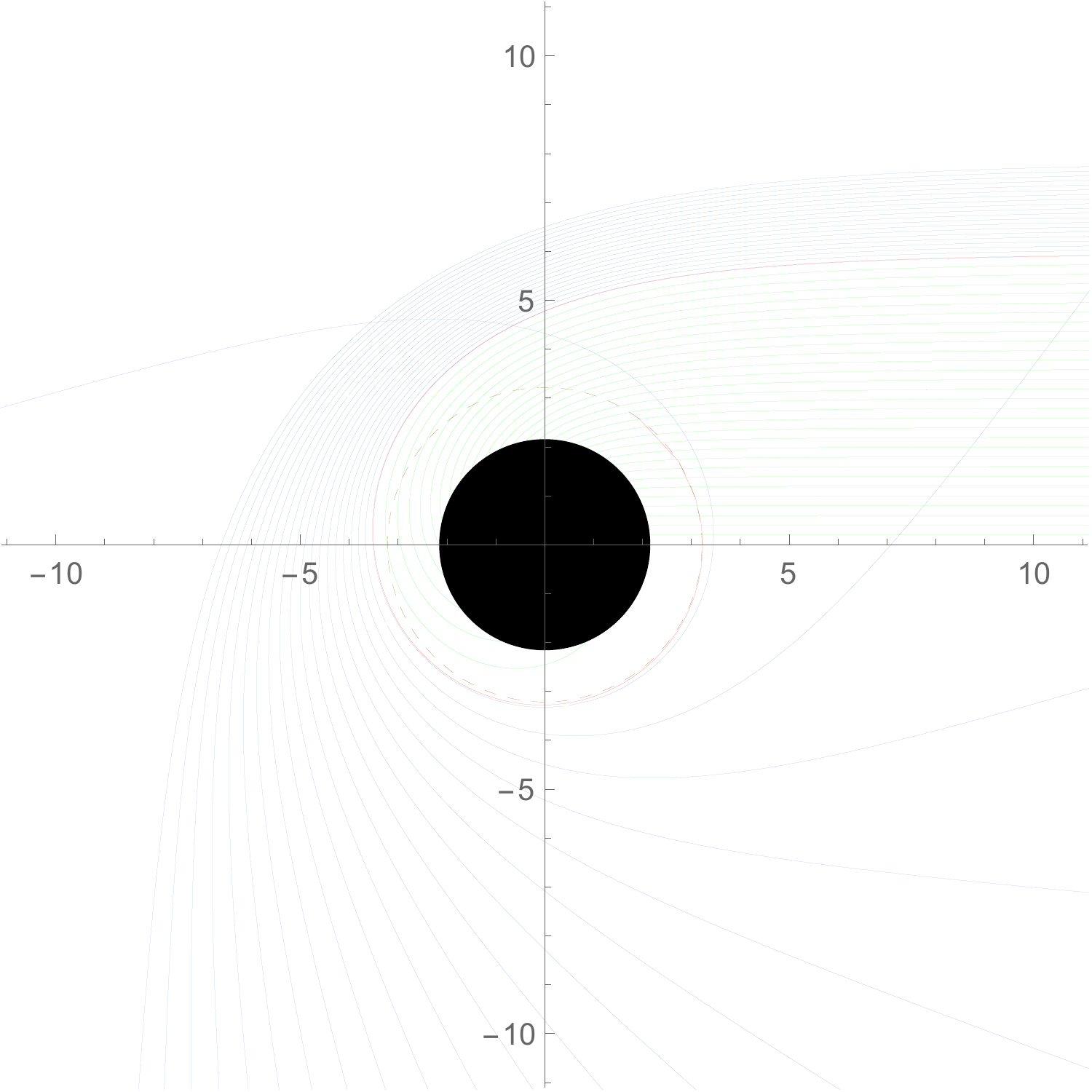}
\hfill
\includegraphics[width=.45\textwidth]{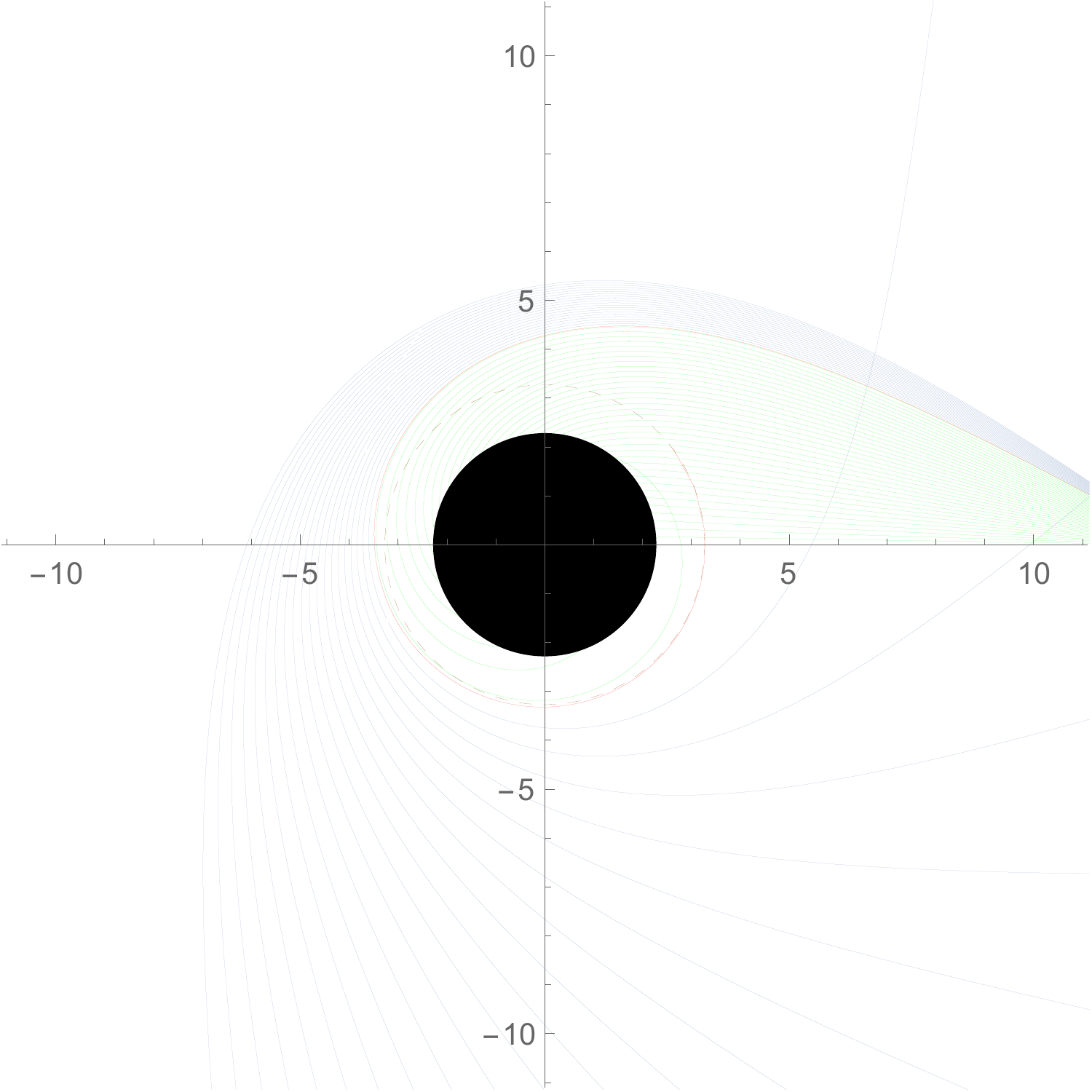}
\caption{\label{fig2}  The trajectory of the light ray for $w=-0.5$ (left) and $w=-0.7$ (right) with $M=1$, $a=0.05$. They are shown in the polar coordinates $(r, \psi)$.  The red lines, green lines and blue lines correspond to $b=b_{ph}$,  $b<b_{ph}$ and $b>b_{ph}$ respectively. The spacing of impact parameter is $\Delta b=1/5$ for each light ray. Black hole is
shown as the solid black disk and the photon orbit as a dashed  line.}
\end{figure}

By virtue of the Eq.\eqref{gu}, we can solve the  trajectories of the light rays, shown in Figure \ref{fig2}. All the light rays approach the black hole from the right side. The green, red and blue  lines correspond to $b<b_{ph}$, $b=b_{ph}$ and  $b>b_{ph}$ respectively.
Specifically, for the case of $b<b_{ph}$, the light rays (green lines) drop all the way into the black hole. For $b>b_{ph}$, the light rays (blue lines) are deflected but never enter the black hole. In particular, the light rays close to the black hole can even be reflected back to the right side, i.e., the side they approach the black hole.  For $b=b_{ph}$, the light rays (red lines) revolve around the black hole.  This conclusion is consistent with those from the analysis of effective potential in Figure \ref{fig1}. Indeed, region 1, 2, 3 in  Figure \ref{fig1} correspond to the blue, red and green lines in  Figure \ref{fig2}, respectively.

We need to emphasize that due to the existence of the cosmological horizon, different $w$'s will result in different geometries of the light rays. We have assumed that the observer is in the domain of the outer communications and sitting close to the cosmological horizon. Therefore, for the case of $w=-0.5$ (left panel in Figure \ref{fig2}), the cosmological horizon is much far from the event horizon (see Table \ref{tab11} ).
Therefore, the observed entering light rays are approximately parallel, which can be seen from the left panel of Figure \ref{fig2}. However, for the case of $w=-0.7$, the cosmological horizon is close to the event horizon (see Table \ref{tab11}). Therefore, the observed entering light rays behave significantly different from those of $w=-0.5$ (see the right panel of Figure \ref{fig2}), i.e., the entering light rays for $w=-0.7$ are not parallel.

As the light is deflected, a shadow of the black hole will be formed. The light is emitted  by the accretion matters, the profiles of the accretion matters thus are important for the shadows of the black holes. Next, we will investigate the effect of the profiles of the accretions on the shadows of the black holes. We will take the thin disc accretion and spherical accretion as two examples.


\section{Shadows, photon rings and lensing rings  with thin disk accretion }
\label{disk}

In this section, we consider an optically and geometrically thin disk accretion around the black hole, viewed face-on. From \cite{Gralla:2019xty}, we know that an important feature of thin disk accretion is that there are photon rings and lensing rings surround the black hole shadow. The lensing ring  consists of light rays that intersect the
plane of the disk twice outside the horizon, while the photon ring \footnote{In our paper, the definition of photon ring  is different from photon sphere discussed in the later section. In references such as \cite{Gralla:2019xty}, the photon sphere is also called `photon ring'. We distinguish them in our paper.   } comprises light rays that intersect the plane of the disk three or more times.  Thus, the trajectory of the photon is important to distinguish the photon ring and lensing ring.

\subsection{Number of orbits of the deflected light trajectories}
\begin{figure}[h]
\centering 
\includegraphics[width=.55\textwidth]{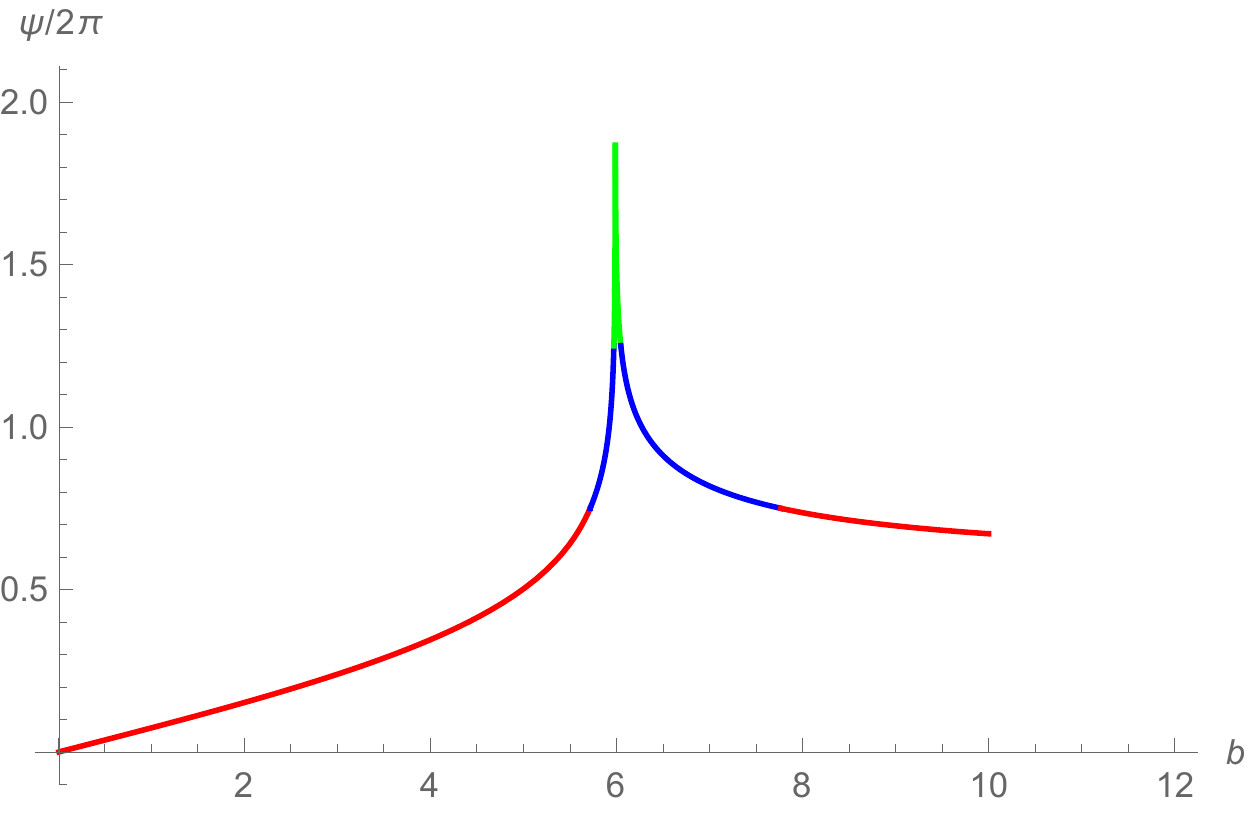}
\hfill
\includegraphics[width=.4\textwidth]{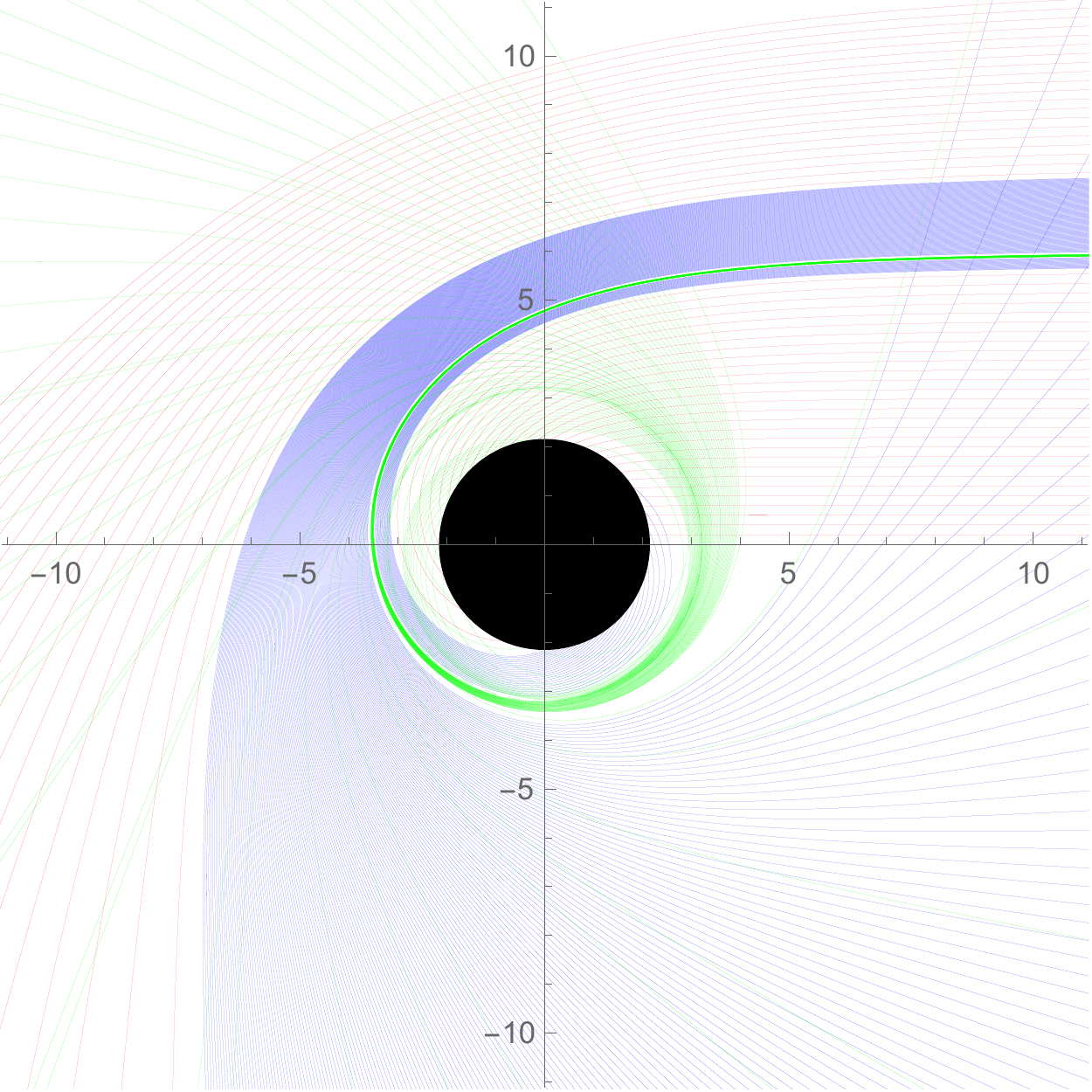}
\includegraphics[width=.55\textwidth]{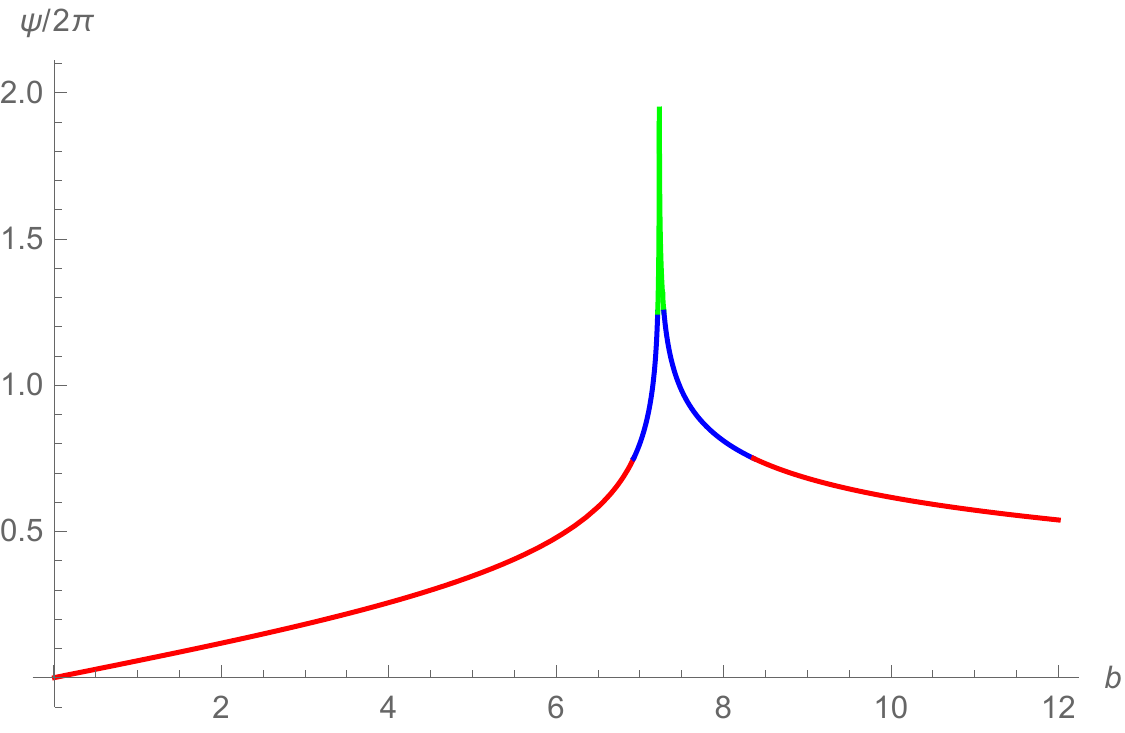}
\hfill
\includegraphics[width=.4\textwidth]{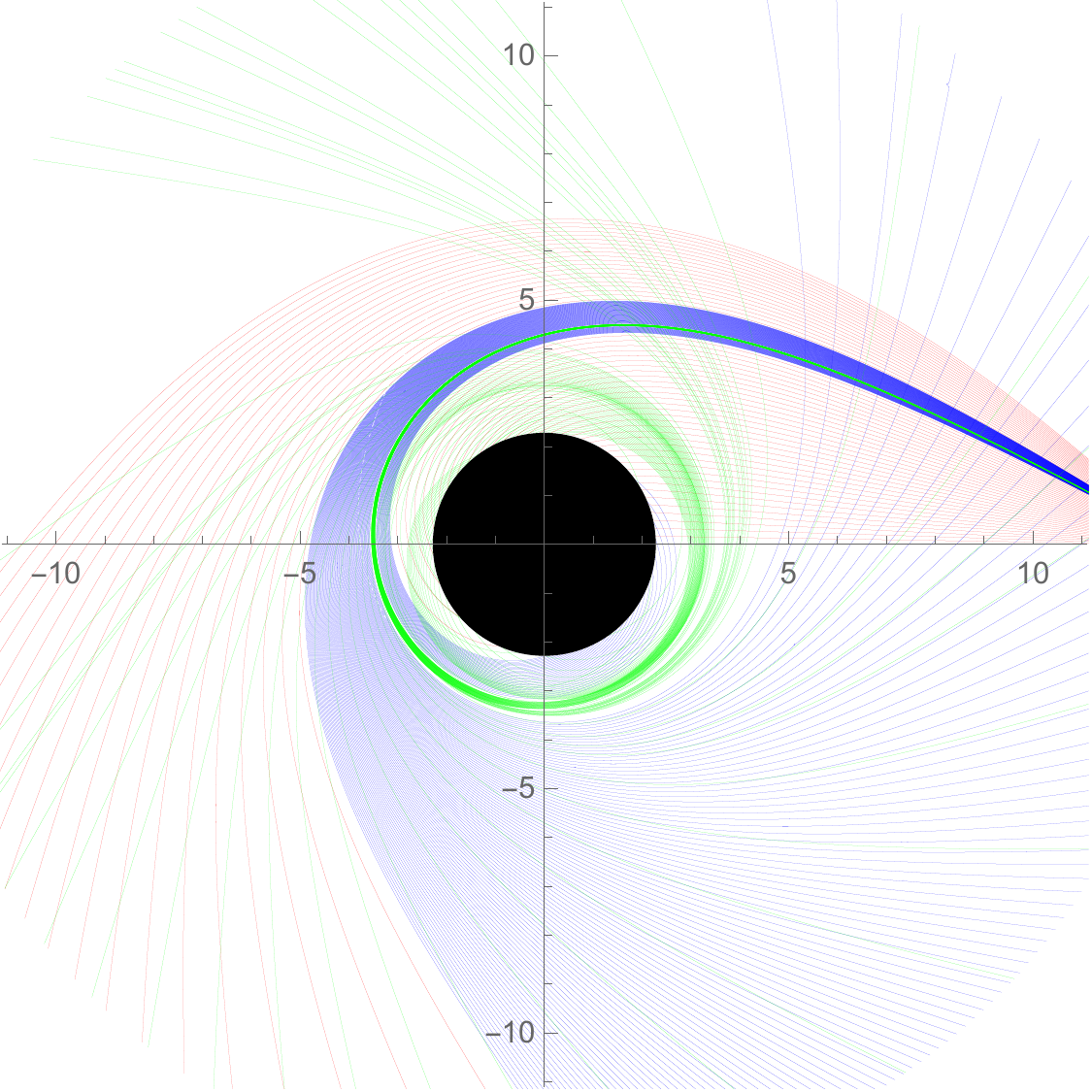}
\caption{\label{fig7} Behavior of photons in the quintessence  black hole as a function of impact parameter $b$ with   $M=1$,  $a=0.05$,  $w=-0.5$ (top row) and $w=-0.7$ (bottom row). (Left column) We show the fractional number of orbits, $n =\psi/2 \pi$, where $\psi$ is the total change in azimuthal angle outside the horizon. The direct, lensing and photon ring correspond to $ n<3/4$ (red), $ 3/4<n<5/4$ (blue), and $ n>5/4$ (green), respectively.  (Right column) We show a selection of associated photon trajectories in the polar coordinates $(b, \psi)$. The spacings in impact parameter are 1/5, 1/100, 1/1000, for the direct (red), lensing (blue), and photon ring (green) bands, respectively. The black holes are shown solid black disks.  }
\end{figure}


The trajectories of the light rays are shown on the right column of Figure \ref{fig7} for $w=-0.5$ (top row) and $w=-0.7$ (bottom row), respectively. In these figures, the red lines, blue lines and green lines  are direct emission, lensing rings and photon rings, respectively. This can be understood from the definition in \cite{Gralla:2019xty} that these light rays intersect the disk plane once, twice and more than twice. It should be noted that due to the domain of outer communications, the light rays going towards the observer will behave differently for different state parameters $w$. For the case of $w=-0.5$, the observer locates near the cosmological horizon which is far away from the event horizon (see Table \ref{tab11}). Thus, the light rays nearby the observer is almost parallel, which is similar to the Schwarzschild black hole \cite{Gralla:2019xty}. However, for $w=-0.7$ the cosmological horizon is near the event horizon (refer to Table \ref{tab11}). Therefore, the light rays going towards the observer are not parallel, which is shown in the bottom row of right panel of Figure \ref{fig7}.

One can also use the total number of orbits, defined by $n=\psi/2 \pi$, to distinguish the trajectories of light rays \cite{Gralla:2019xty}. The total number of orbits are shown in the left column of Figure \ref{fig7} for $w=-0.5$ (top row) and $w=-0.7$ (bottom row), respectively. We still use the red lines, blue lines and green line to represent the direct emission,  lensing rings and photon rings, respectively.  Therefore, from the definitions of these light rings we can explicitly see that the direct emissions correspond to $n<3/4$, while lensing rings correspond to $3/4<n<5/4$ and photon rings correspond to $n>5/4$.  For the case of $w=-0.5$ and  $w=-0.7$, the  parameter intervals of $b$ for the  direct emission, photon rings and lensing rings are listed in  following Eq.(\ref{w5}) and  Eq.(\ref{w7}).
\begin{align} \label{w5}
  w =-0.5
    \begin{cases}
   \text{Direct emission}: n<3/4,~~~
    b < 5.71774  ~~ \text{and} ~~  b> 7.76682 \\
     \text{Lensing ring}: 3/4<n<5/4,~~~
   5.71774<b< 5.97314 ~~  \text{and}   ~~ 6.04784<b< 7.76682  \\
       \text{Photon ring}: n>5/4,~~~
      5.97314< b < 6.04784
    \end{cases}
\end{align}
\begin{align}  \label{w7}
  w =-0.7
    \begin{cases}
   \text{Direct emission}: n<3/4,~~~
    b < 6.92848  ~~ \text{and} ~~  b> 8.36103 \\
     \text{Lensing ring}: 3/4<n<5/4,~~~
  6.92848<b< 7.2156 ~~  \text{and}   ~~ 7.28972<b< 8.36103  \\
       \text{Photon ring}: n>5/4,~~~
      7.2156< b < 7.28972
    \end{cases}
\end{align}

\subsection{Observed specific intensities and transfer functions}

Next, we will investigate the observed specific intensity of the thin disk accretion. We assume that the thin disk emits isotropically in the rest frame of static worldlines. As in \cite{Gralla:2019xty}, we take the disk to lie in the equatorial plane of the black hole, with a static observer at the North pole. We denote the emitted
specific intensity and frequency as $I_{e}(r)$ and $\nu_{e}$, the observed specific intensity and frequency as $I_{obs}(r)$ and $\nu$. Since $I_{e}/\nu_e^3$  is conserved along a light ray from Liouville's theorem, the observed specific intensity thus can be written as
\begin{equation}
 I_{obs}(r)=f(r)^{3/2}I_{e}(r).
\end{equation}
The total specific intensity is obtained by integrating specific intensity with different frequencies
\begin{equation}
 I=\int I_{obs}(r) d\nu=\int f(r)^{2}I_{e}(r)d\nu_e=f(r)^{2}I_{em}(r),
\end{equation}
where $I_{em}(r)=\int I_{e}(r)d\nu_e$ is the total  emitted specific intensity near the accretion.

If a light ray traced backwards from the observer intersects the disk, it will pick up brightness from the disk emission. For $3/4<n<5/4$, the light ray will bend around the black hole and hit the opposite side of the disk from the back (see the blue lines in Figure \ref{fig7}). Thus, it will pick up additional brightness from this second passage through the disk. For $n>5/4$, the light ray will bend around the black hole more, and then hit the front side of the disk once again (see the green lines in Figure \ref{fig7}). This leads to additional brightness from the third passage through the disk.  Therefore, the observed intensity is a sum of the intensities from each
intersection, that is
\begin{equation}
 I(r)=  \sum\limits_{m}  f(r)^{2}I_{em}\mid_{r=r_m(b)}, \label{lir}
\end{equation}
where $r_{m(b)}$, called transfer function, is the radial position of the $m^{th}$ intersection
with the disk plane outside the horizon. For simplicity, we have neglected absorption of the light for the thin accretion, which may decrease the observed intensity from the additional passages.

\begin{figure}[tbp]
\centering 
\includegraphics[width=.45\textwidth]{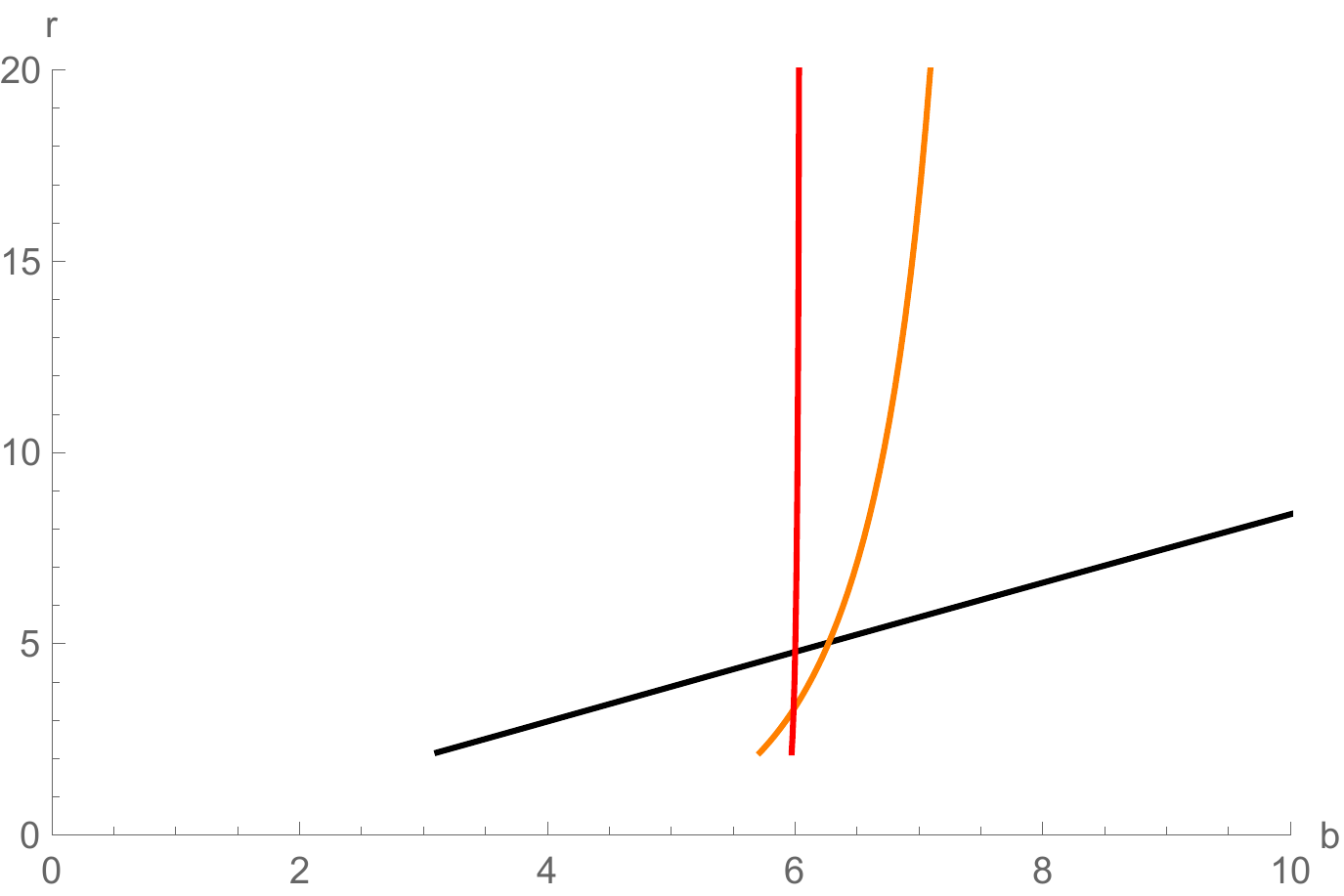}
\hfill
\includegraphics[width=.45\textwidth]{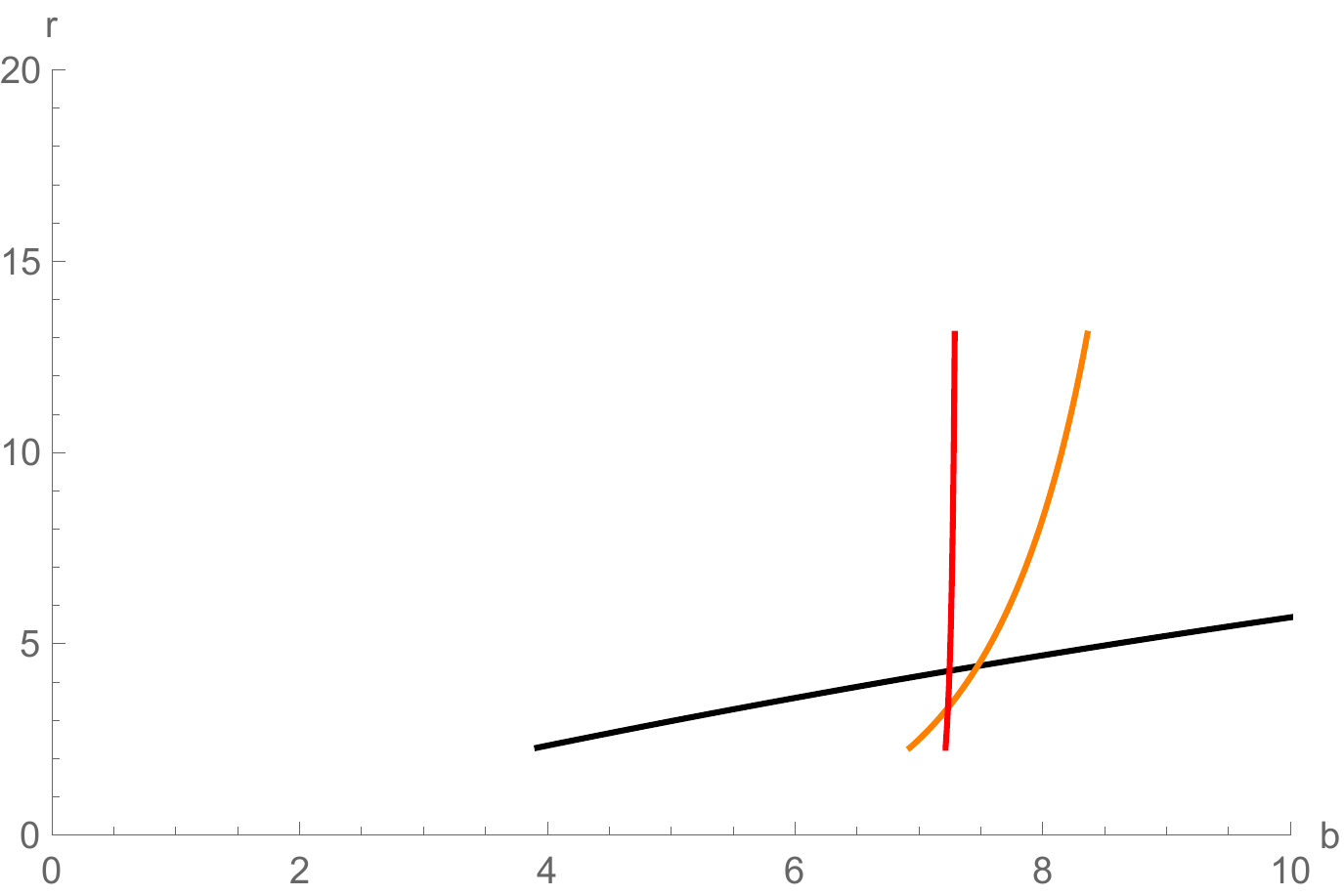}
\caption{\label{fig9} The first three transfer functions $r_m(b)$ for the thin disk in the quintessence black hole with $w=-0.5$ (left panel) and $w=-0.7$ (right panel). The black lines, orange lines and red lines stand for  the radial coordinate of the first, second, and third intersections with a face-on thin disk outside the horizon. }
\end{figure}

The transfer function describes the relation between the radial coordinate $r$ and the impact parameter $b$. The slope of the transfer function, $dr/db$, is the demagnification factor. For different state parameters $w$, the  transfer functions with respect to impact parameter $b$ are shown in Figure \ref{fig9}.  The black line, orange line and red line represent the first ($m=1$), second ($m=2$) and third ($m=3$) transfer functions. The first   transfer function corresponds to the direct image of the disk, and its slope is about 0.85 (0.67) for $w=-0.5$ ($w=-0.7$). Therefore, the direct image profile is  actually the redshifted source profile. The second
  transfer function  corresponds to the lensing ring (including  the photon ring). In this case, the observer
will see a highly demagnified image of the back side of the disk. Finally, the third transfer function corresponds to the photon ring. In this case, one will see an extremely demagnified image of the front side of the
disk since the slope is about infinite. It thus  contribute negligibly to the total brightness of the image.

\subsection{Shadows, lensing rings and photon rings}
\begin{figure}[h]
\centering 
\includegraphics[width=.35\textwidth]{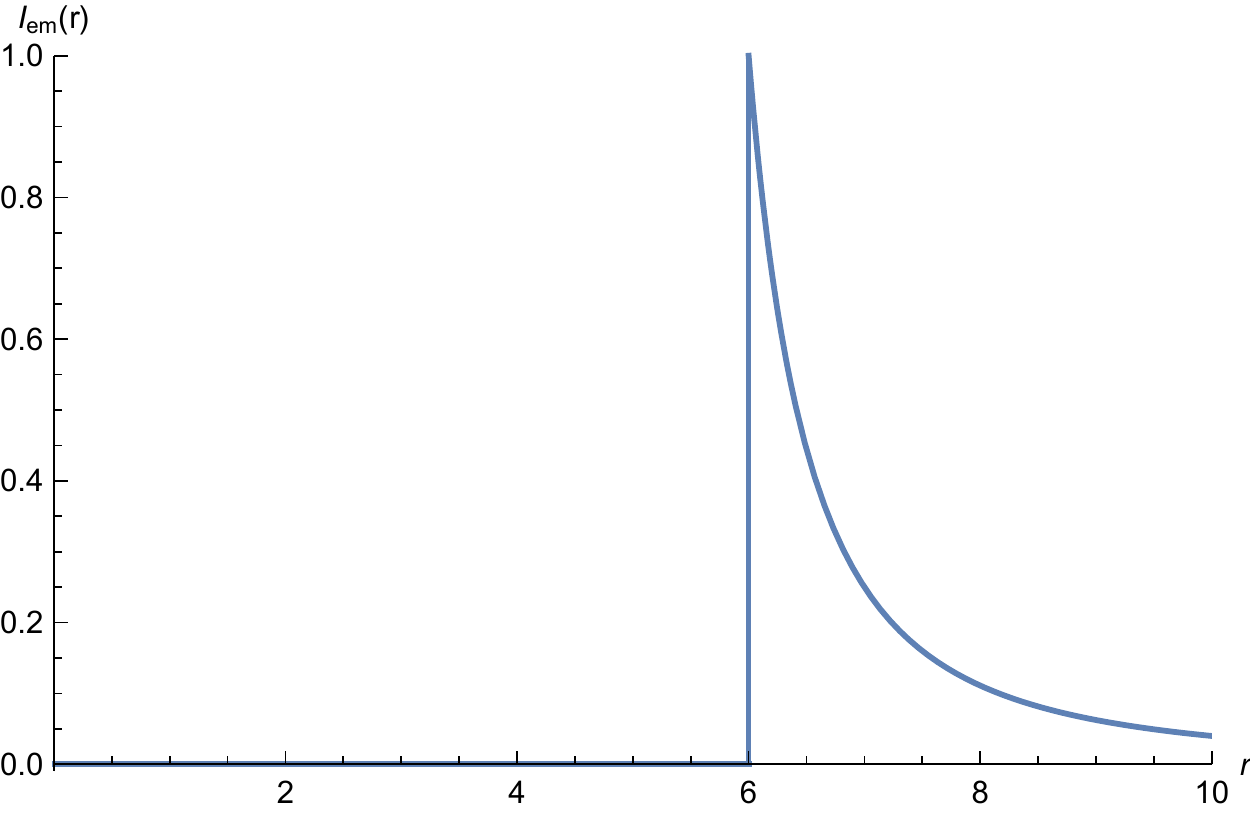}
\includegraphics[width=.35\textwidth]{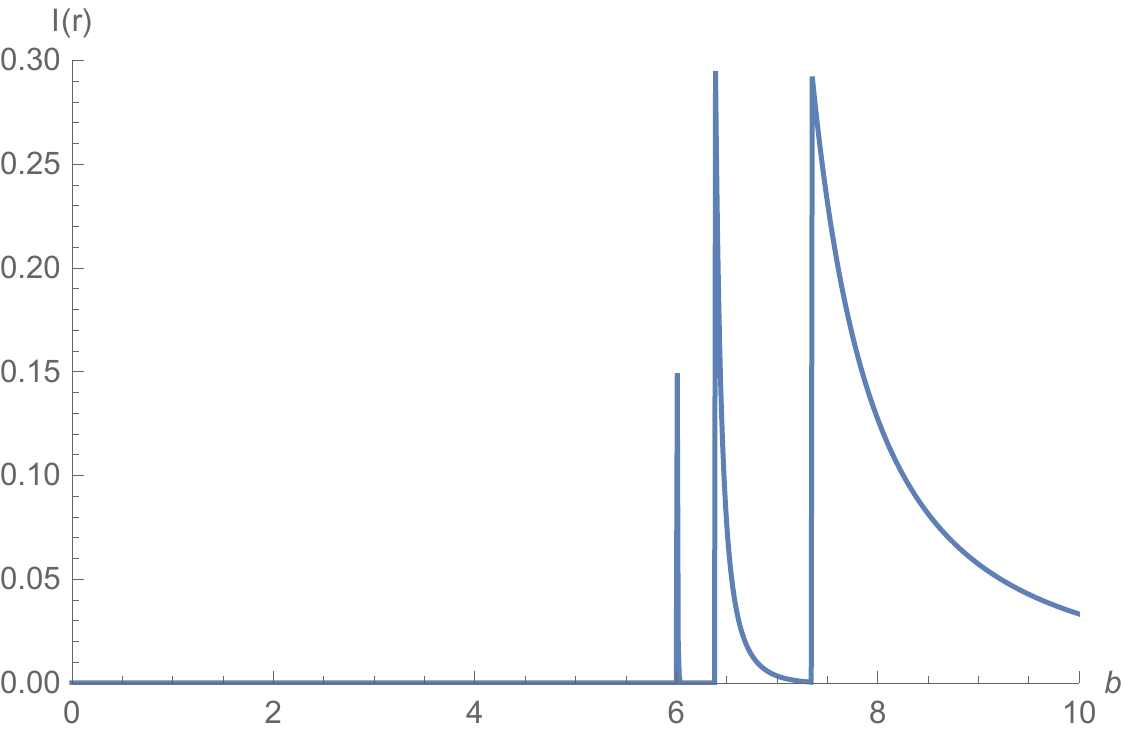}
\includegraphics[width=.28\textwidth]{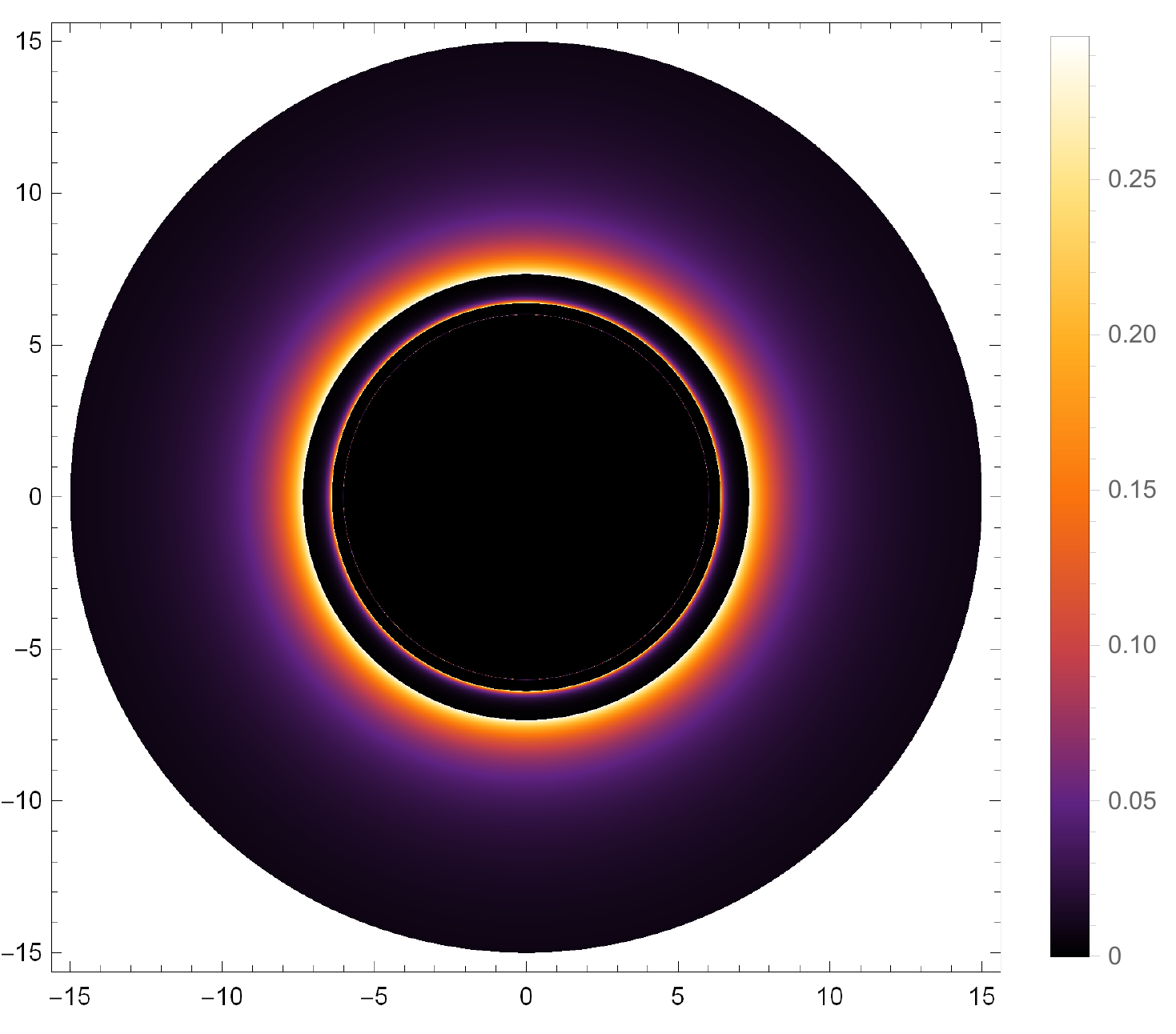}
\includegraphics[width=.35\textwidth]{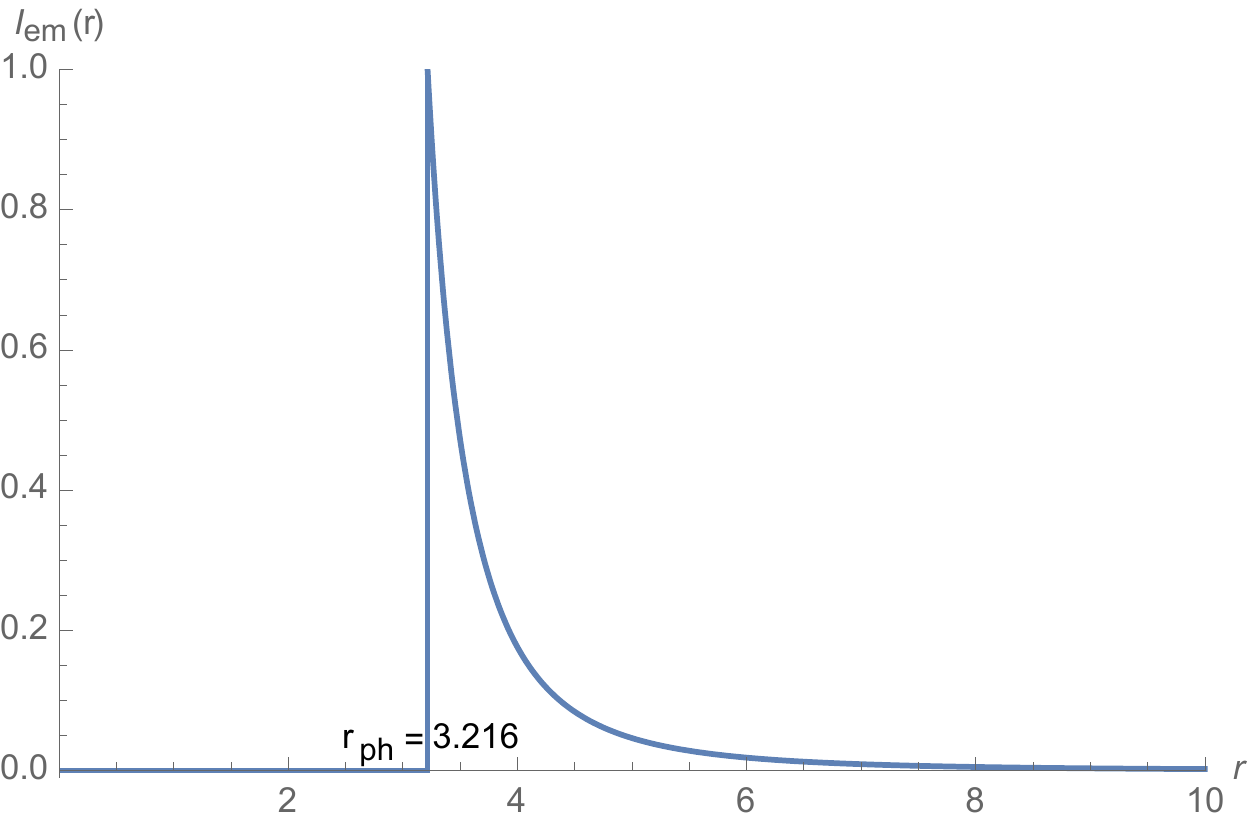}
\includegraphics[width=.35\textwidth]{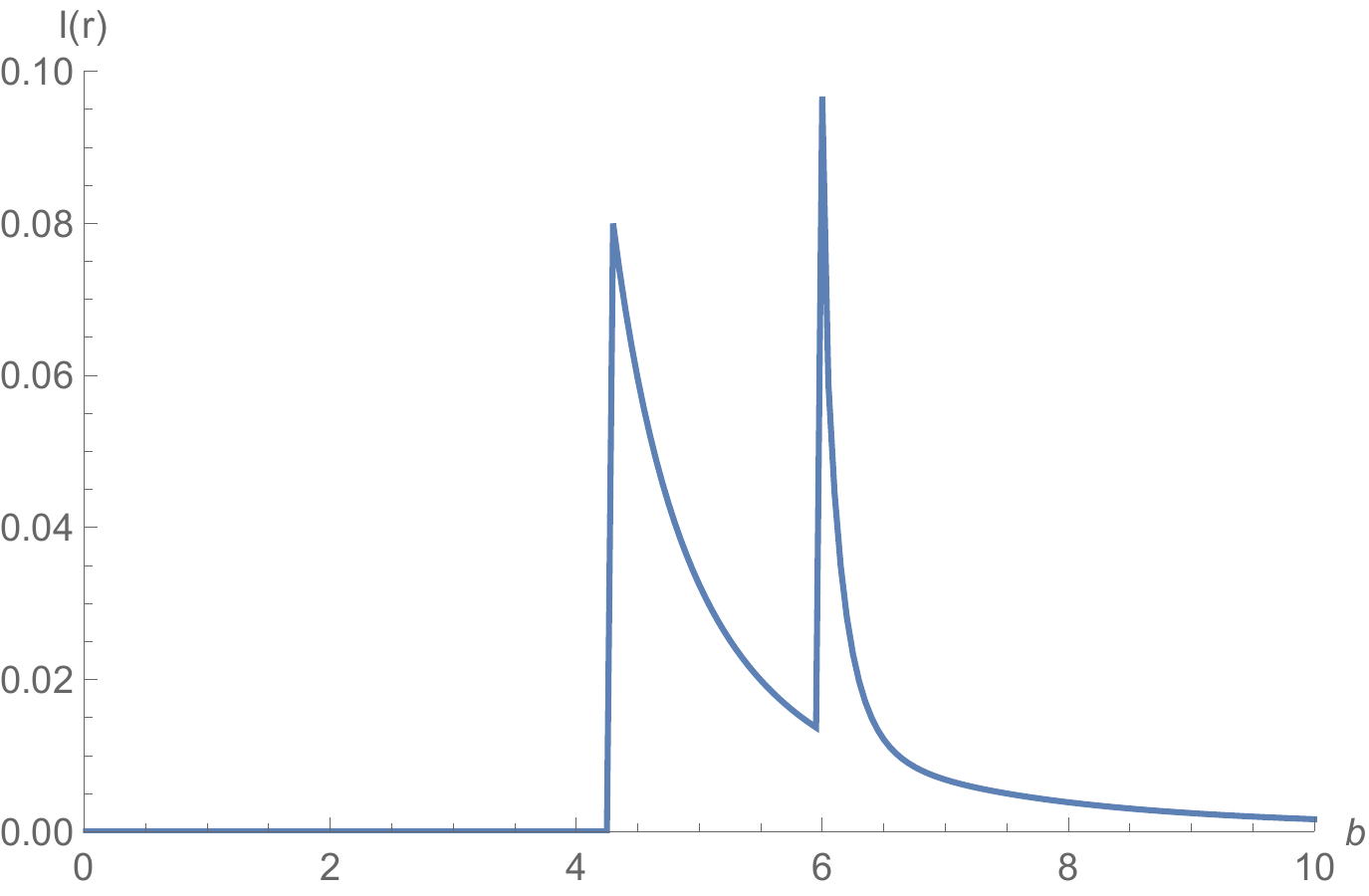}
\includegraphics[width=.28\textwidth]{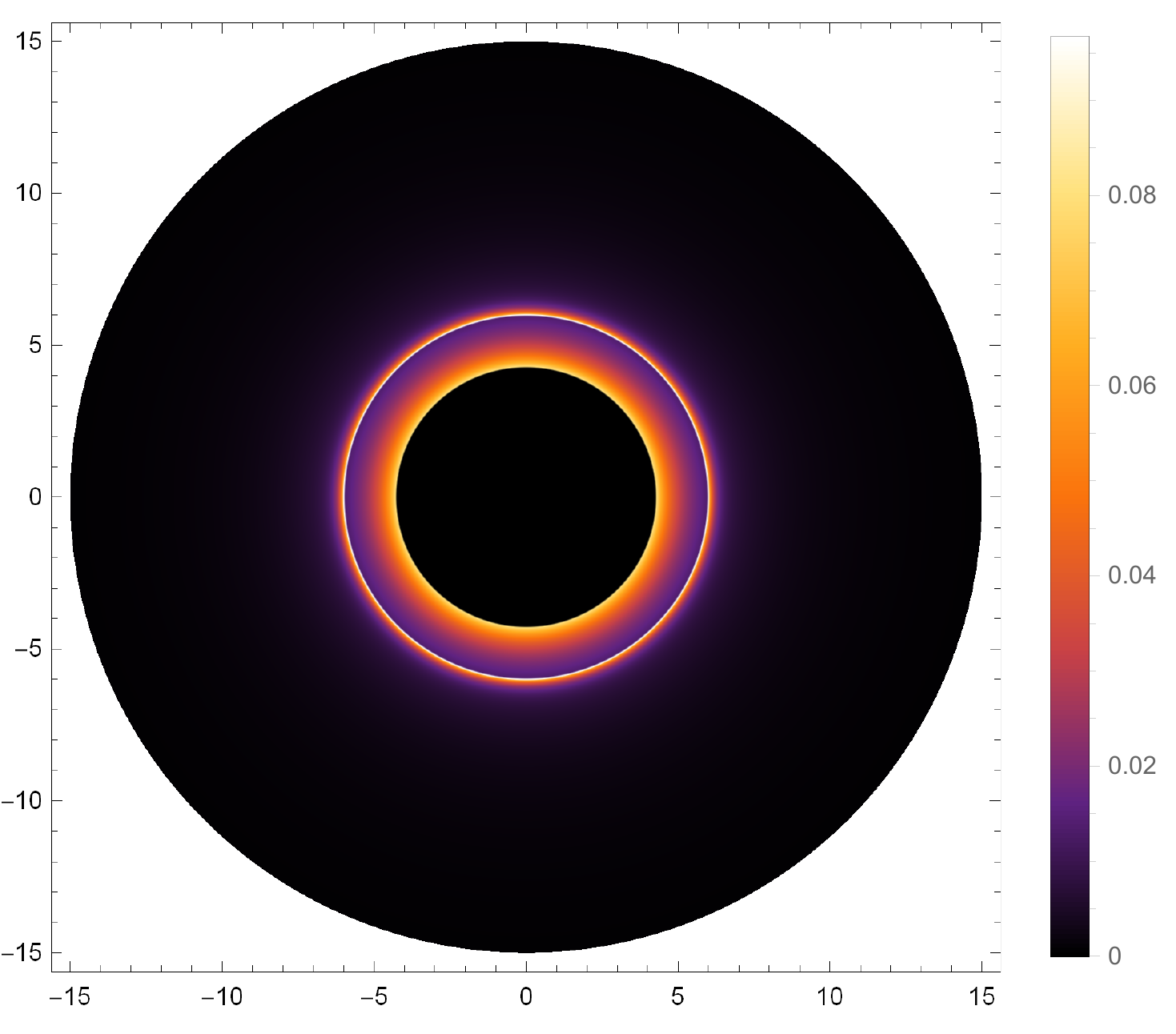}
\includegraphics[width=.35\textwidth]{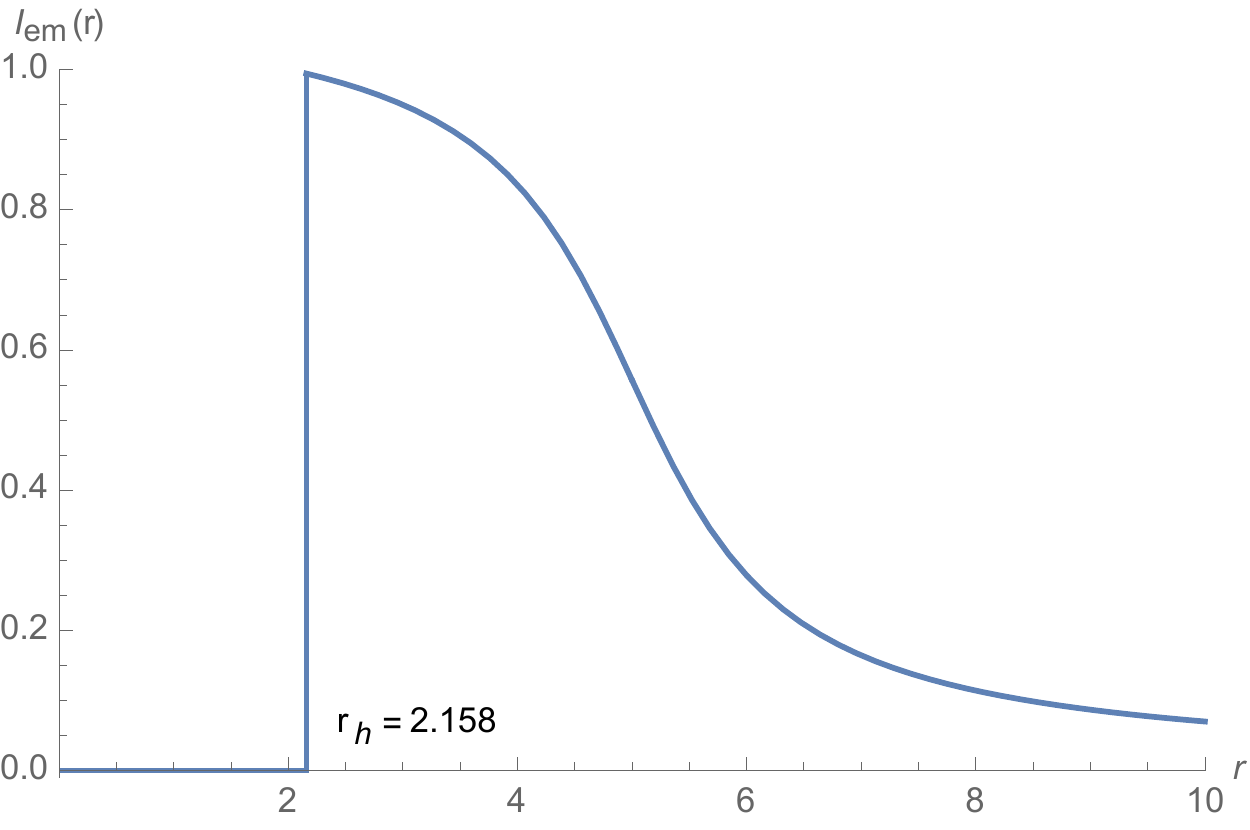}
\includegraphics[width=.35\textwidth]{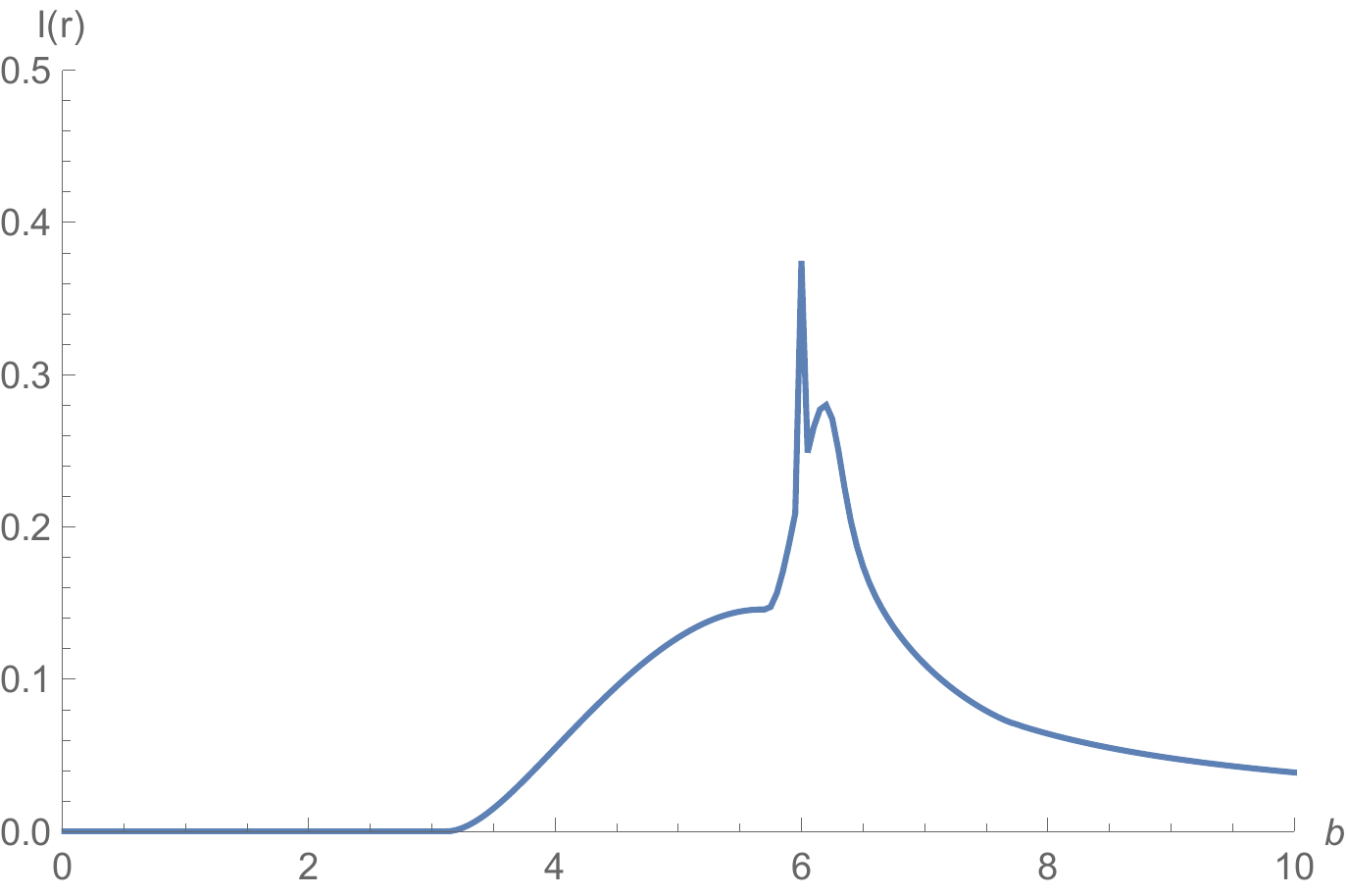}
\includegraphics[width=.28\textwidth]{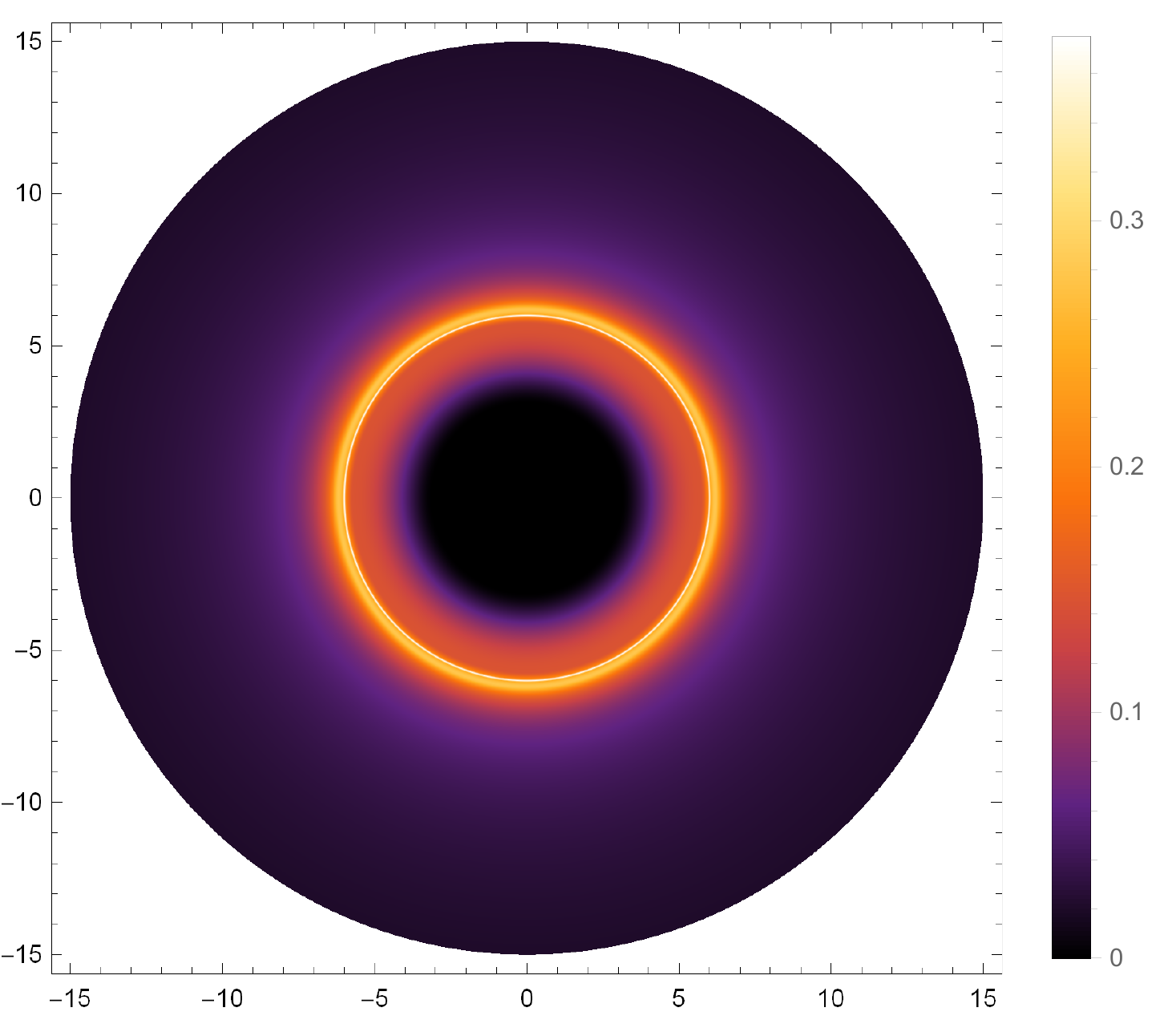}
\caption{\label{fig100}  Observational appearances of a geometrically and optically thin disk with different profiles near black hole with $M=1$,  $a=0.05$,  $w=-0.5$, viewed from a face-on orientation. The left column shows the profiles of various emissions $I_{em}(r)$. The middle column exhibits the observed emission $I(r)$ as a function of the impact parameter $b$.  The right column shows the 2-dim density plots of the observed emission $I(r)$. The details can be found in the main texts. 
}
\end{figure}

\begin{figure}[tbp]
\centering 
\includegraphics[width=.35\textwidth]{fig1911.pdf}
\includegraphics[width=.35\textwidth]{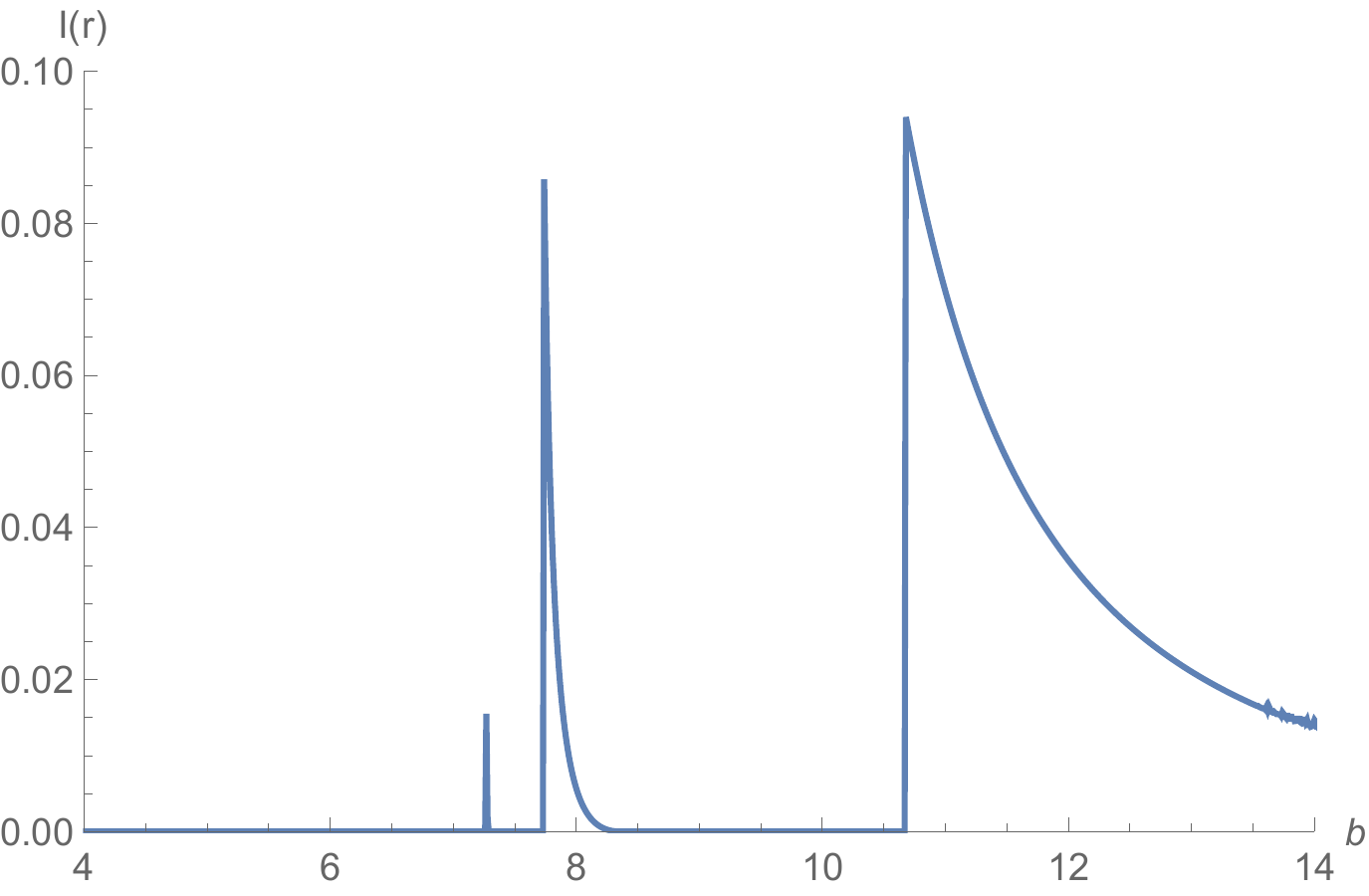}
\includegraphics[width=.28\textwidth]{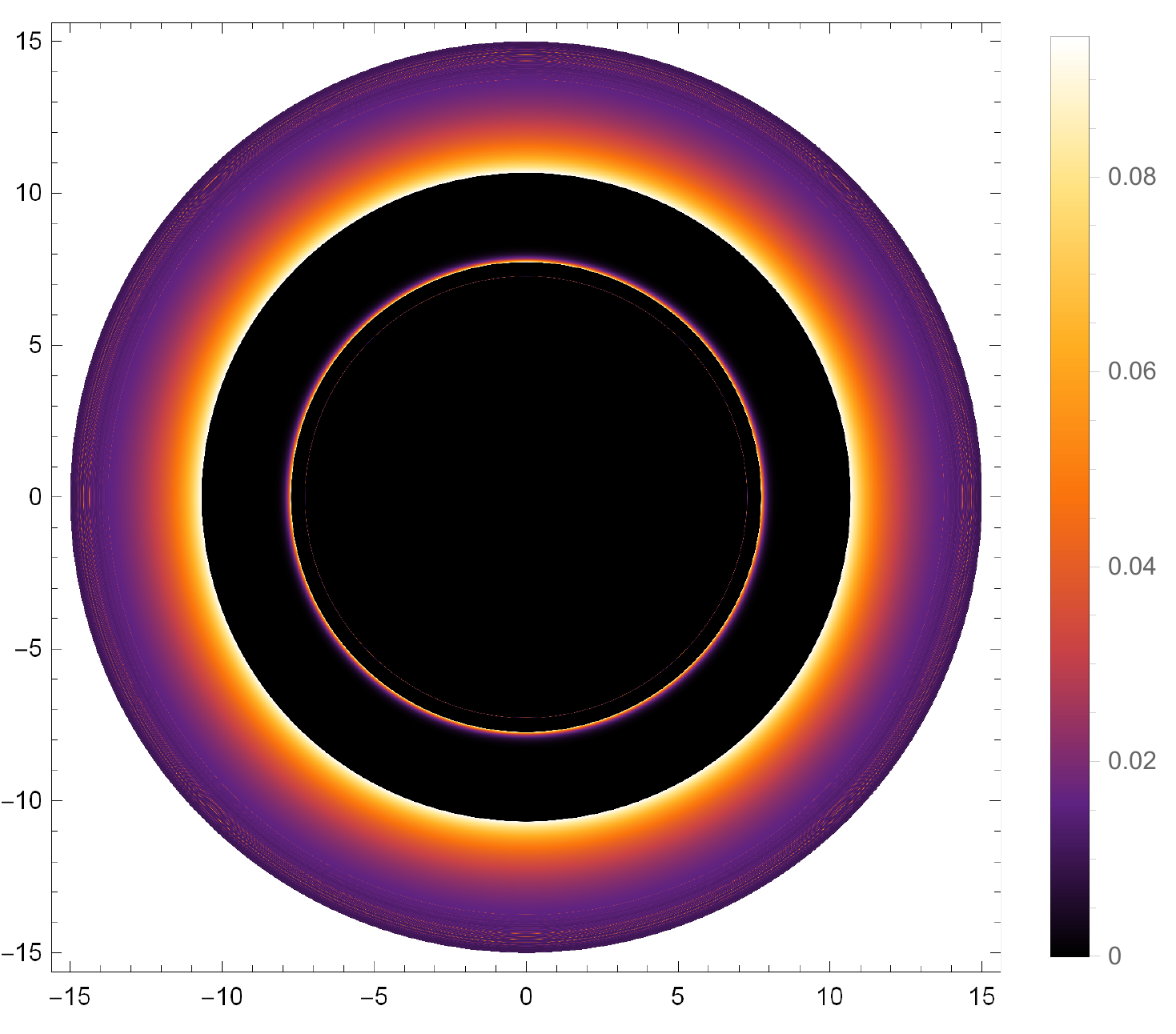}
\includegraphics[width=.35\textwidth]{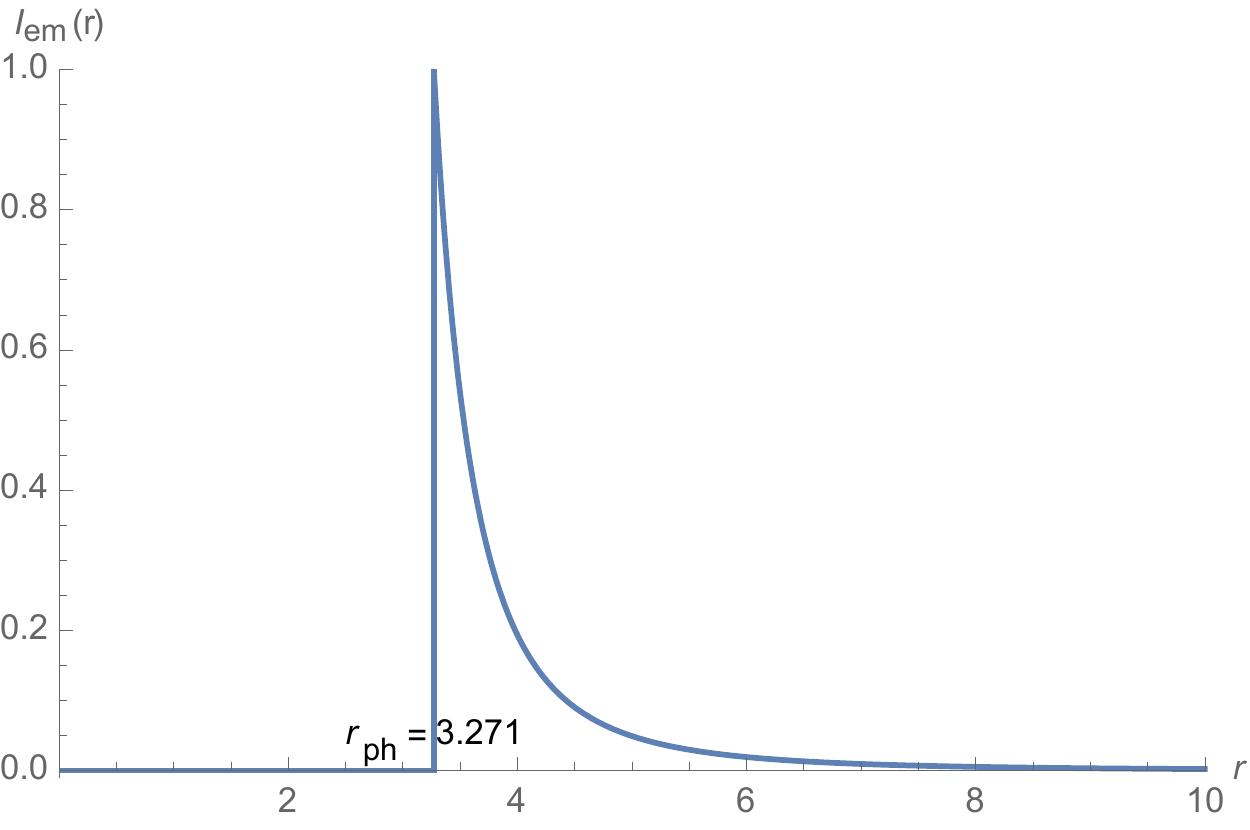}
\includegraphics[width=.35\textwidth]{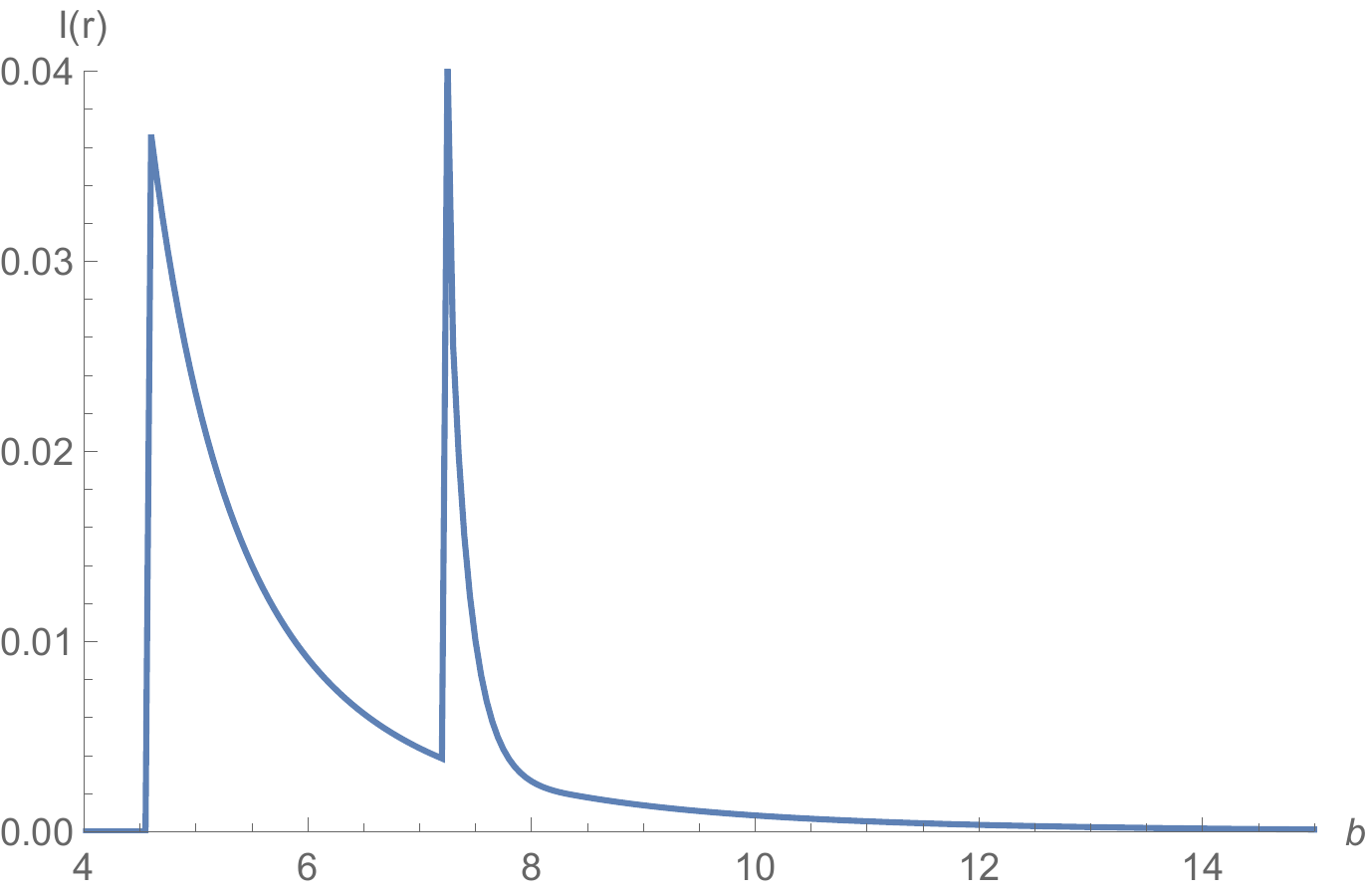}
\includegraphics[width=.28\textwidth]{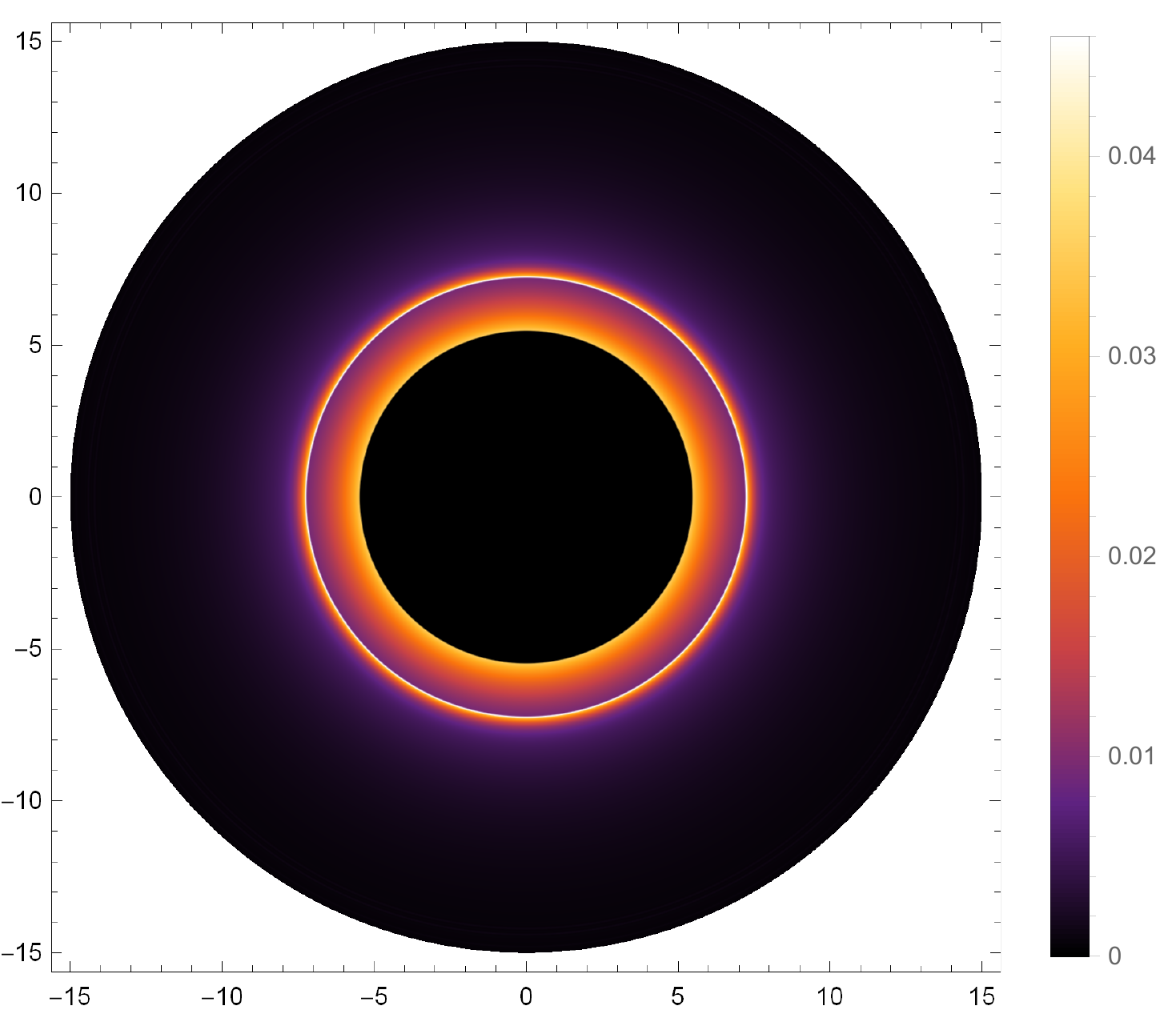}
\includegraphics[width=.35\textwidth]{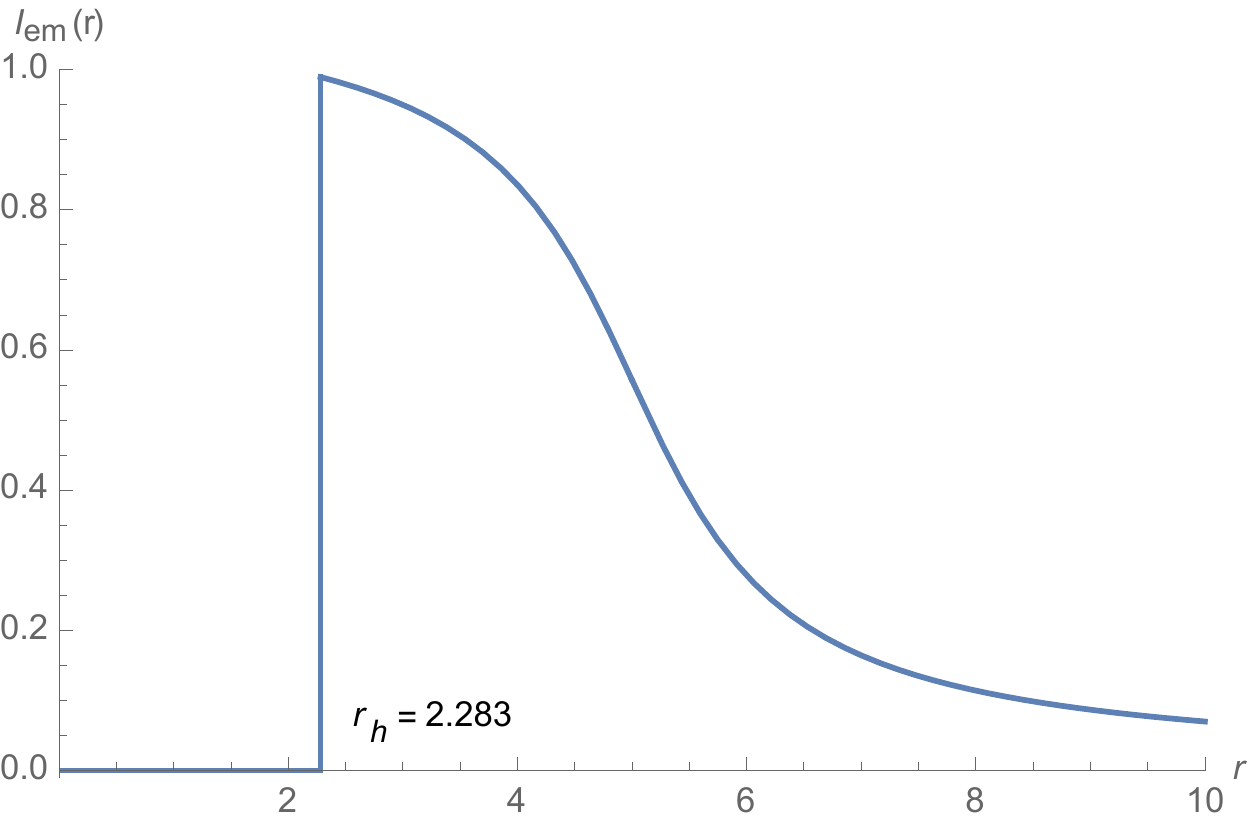}
\includegraphics[width=.35\textwidth]{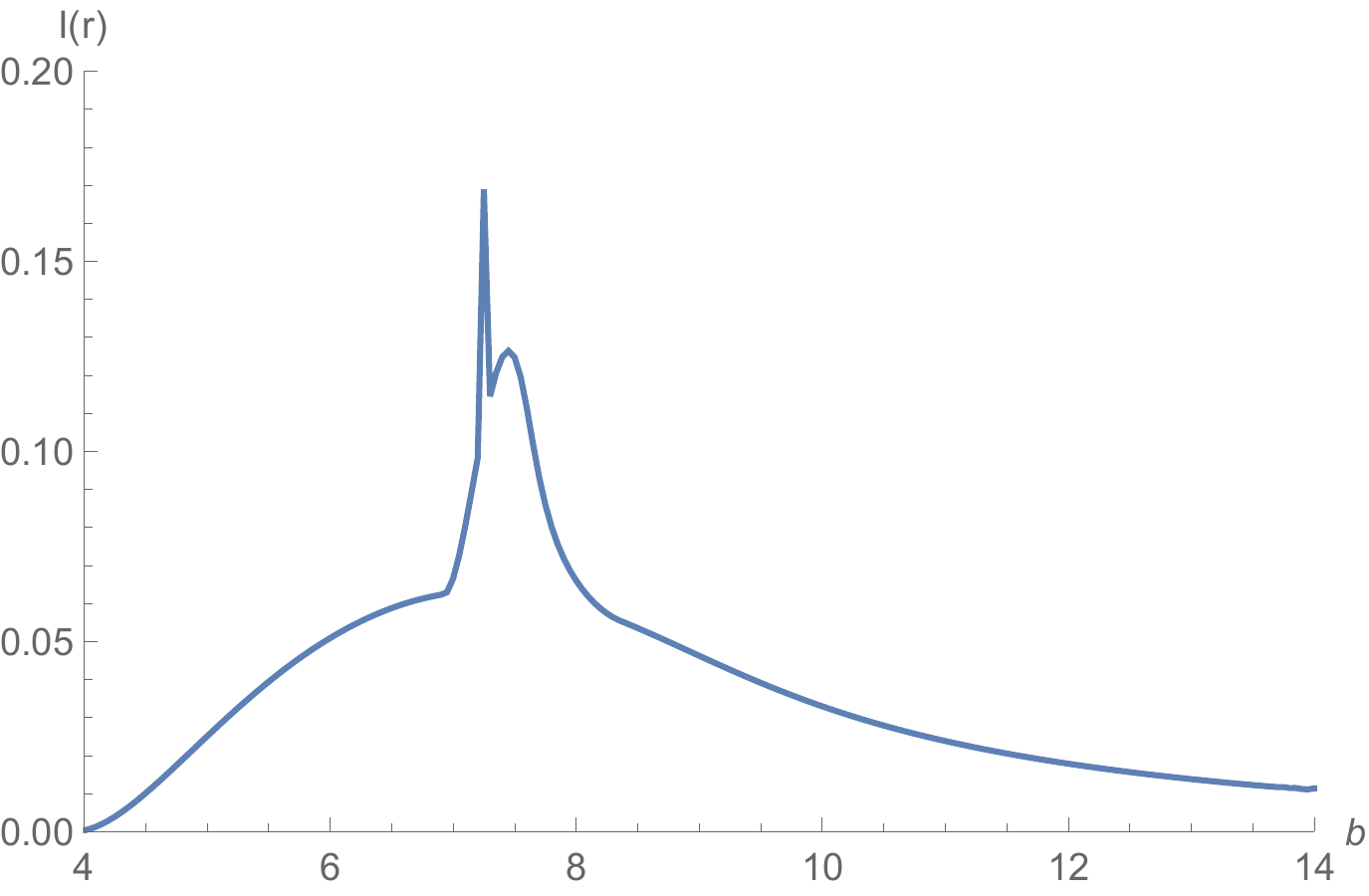}
\includegraphics[width=.28\textwidth]{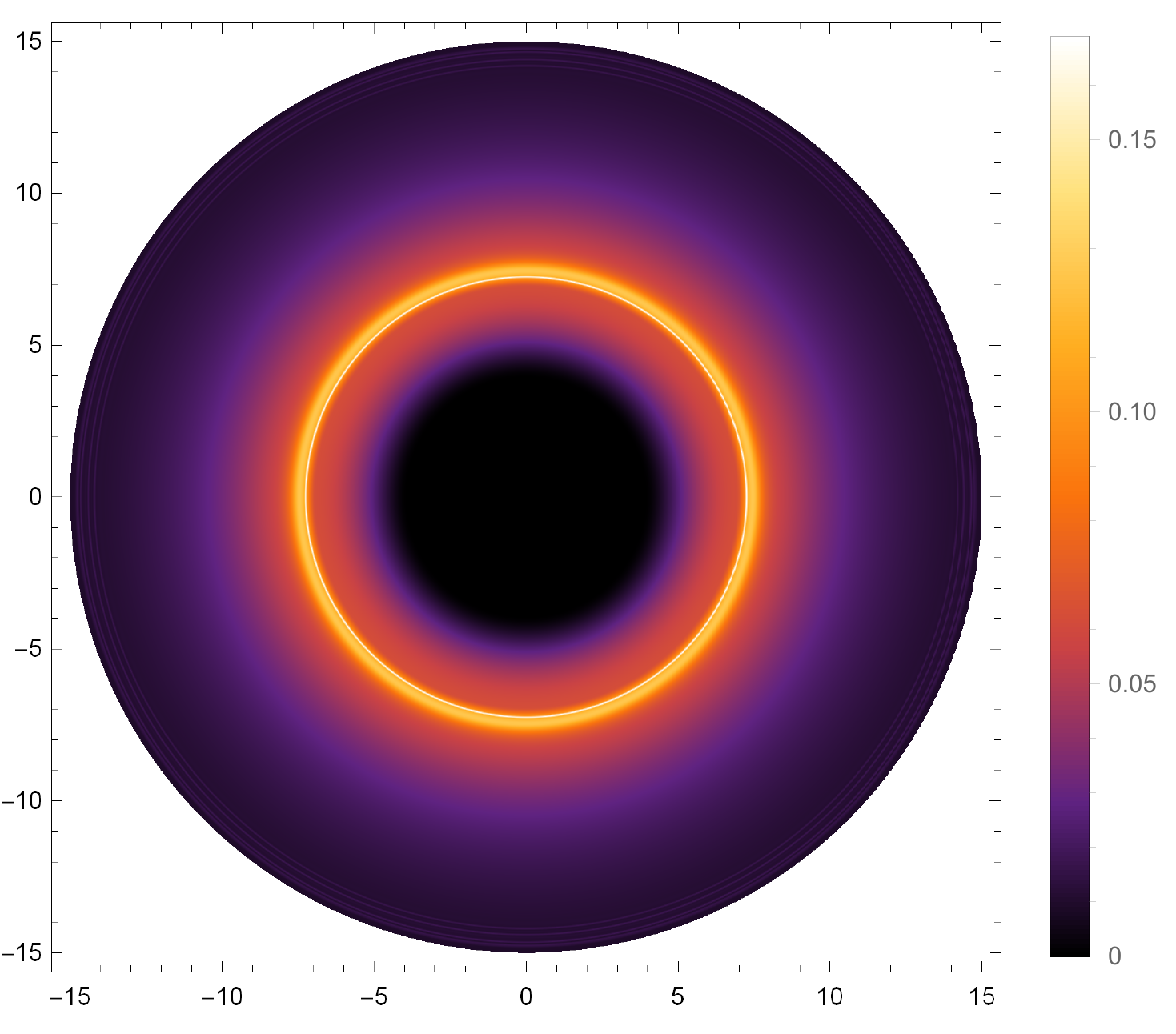}
\caption{\label{fig11}  Observational appearances of a geometrically and optically thin disk with different profiles near black hole with $M=1$,  $a=0.05$,  $w=-0.7$, viewed from a face-on orientation. The left column shows the profiles of various emissions $I_{em}(r)$. The middle column exhibits the observed emission $I(r)$ as a function of the impact parameter $b$.  The right column shows the 2-dim density plots of the observed emission $I(r)$. The profiles are similar to those of $w=-0.5$ in Figure \ref{fig100}. }
\end{figure}

Having obtained the transfer functions, we can further obtain the specific intensity on the basis of  Eq.(\ref{lir}) as the emitted specific intensity is given. In this paper, we assume the following toy-model  functions which may emit by some thin matters. Firstly, we assume that $I_{em}(r)$ is a decay function of the power of second order,
\begin{align}
    I_{em}(r) =\begin{cases}\left(\frac{1}{r-5}\right)^2, &  r> 6  \\
    0, &r \leq 6. \label{fun1}
    \end{cases}
\end{align}
This profile is plotted in the top left panels in Figure \ref{fig100} for $w=-0.5$ and Figure \ref{fig11} for $w=-0.7$. This emission is sharply peaked at $r = 6M$, and then drops to zero abruptly as $r<6M$. In this example, the region of emission is well outside the photon sphere (see Table \ref{tab11}).
Secondly, we assume that the emission is a decay function of the power of third order as
\begin{align}
    I_{em}(r) =\begin{cases}\left(\frac{1}{r-(r_{ph}-1)}\right)^3, &  r> r_{ph}   \\
    0, &r \leq r_{ph} \label{fun2}
    \end{cases}
\end{align}
But this time the peak of the mission is at the photon sphere $r_{ph}$. The profiles of this emission can be seen from the middle left panels in Figure \ref{fig100} for $w=-0.5$ and Figure \ref{fig11} for $w=-0.7$.
Finally, we assume a moderate decay of emission as
\begin{align}
    I_{em}(r) =\begin{cases} \frac{\frac{\pi }{2}-\tan ^{-1}(r-5)}{\frac{\pi }{2}-\tan ^{-1}(-3)}
    , &  r>r_h  \\
    0,  &r \leq r_h  \label{fun3}
    \end{cases}
\end{align}
This emission is outside of the horizon $r_{h}$, which is shown in the bottom left panels in Figure \ref{fig100} for $w=-0.5$ and Figure \ref{fig11} for $w=-0.7$.


For the case of $w=-0.5$, the observed specific intensities are shown in the middle and right columns in Figure \ref{fig100}. The middle column is about the one dimensional functions of $I(r)$ with respect to $b$, while the right column is about the two dimensional density plots of $I(r)$, viewed faced-on.
In the first row the emission from the accretion is well outside the photon sphere (left panel). The corresponding observed direct image decays similarly as $b>7.5M$ due to the gravitational lensing (middle panel). However, the observed lensing ring emission is confined to a thin region $6.4M<b<7.4M$, since the back side image is highly demagnified for $r>6M$. The photon ring emission is the spike at $b\sim 6M$, which is hardly to see in the right panel of the density plots. One needs to zoom in the plot to see the photon ring. Therefore, from the observer, the lensing ring makes a small contribution to the total brightness while the photon ring makes negligible contributions.


In the second row of Figure \ref{fig100}, the emission starts from the photon sphere $r_{ph}=3.216$. The most important difference from the first row is that the observed lensing ring and photon ring emission are now superimposed on the direct image emission, which decays from $b>4.25M$. The lensing ring has a spike in the brightness in the range of $6M<b<6.5M$, while the photon ring has an even narrower spike at $b\sim6M$ which is hardly to be distinguished from the lensing ring.  In this case, the lensing ring still makes a very small contribution to the total brightness  and the photon ring  makes a negligible contribution.

Finally, in the third row of Figure \ref{fig100}, the emission extends to the event horizon $r_h=2.158$ and the decay of the emission is much moderate compared to the first and second row. The inner edge of the observed intensity at $b\sim 3.2M$ is the lensed position of the event horizon. The increase of the intensity outside of the central dark area is due to the gravitational redshift. The very narrow spike at $b\sim6M$ is the photon ring,
while the broader bump at $b\sim6.2M$ is the lensing ring. In this case, the lensing ring makes a prominent brightness to the observed intensity, but the photon ring is still entirely negligible.

From Figure \ref{fig100}, we see that the dark central regions are different for different profiles of the emissions. But the locations of the photon ring always stay at $b\sim 6M$.  In addition, for all the cases, we find that the observed brightness mainly stems from the direct emission, while the lensing ring provides only a small contribution, and the photon sphere provides a negligible  contribution.

For the case of $w=-0.7$, we plot the profiles of the emission and observed intensities in Figure \ref{fig11}. They behave similarly to those of $w=-0.5$.  The differences are the quantities of the intensities, as well as the locations of photon ring and lensing ring.

\section{Shadows and photon spheres with spherical accretions }
\label{spherical}

 In this section, we are going to investigate the shadows and photon spheres with the spherical accretions for various state parameters $w$.

\subsection{Shadows and photon spheres with a static spherical accretion }

In this section, we intend to investigate the shadow  and photon sphere  of a   quintessence black hole with a static spherical accretion.
We mainly focus on the
specific intensity
observed by the observer $\rm (erg s^{-1} cm^{-2} str^{-1} Hz^{-1})$, which can be expressed as
\cite{Jaroszynski:1997bw, Bambi:2013nla}
\begin{equation} \label{intensity}
I = \int_\gamma g^3 j (\nu_{\rm e}) dl_{\rm p} ,
\end{equation}
in which  $g = \nu_{\rm o}/\nu_{\rm e}$ is the redshift factor, while $\nu_{\rm  e}$ is the radiated
photon frequency and $\nu_{\rm o}$ is the observed photon frequency, $j (\nu_{ \rm e})$ is the
emissivity per unit volume measured in the static frame of the emitter, $dl_{\rm p}$ is the
infinitesimal proper length, and $\gamma$ stands for the trajectory of the light ray.

In the quintessence black hole \eqref{metric}, the redshift factor  $g =f(r)^{1/2}$.  In addition, we assume that the radiation of light is monochromatic with fixed a frequency $\nu_s$, thus the specific emissivity takes the form as
\begin{equation}
j (\nu_{\rm e})\propto \frac{\delta(\nu_e-\nu_s)}{r^2},  \label{profile}
\end{equation}
 where we have adopted the $1/r^2$ profile as in \cite{Bambi:2013nla}.  According to  Eq.(\ref{metric}), the  proper length measured in the rest frame of the emitter is
\begin{eqnarray}
dl_{\rm p}&=&\sqrt{f(r)^{-1}dr^2+r^2 d \psi^2} \nonumber \\
  &=&\sqrt{f(r)^{-1}+r^2 \left(\frac {d \psi}{dr}\right)^2}~ dr, \label{dl}
\end{eqnarray}
in which $ d \psi/dr$ is given by the inverse of Eq.(\ref{drp}). In this case, the specific intensity observed by the static observer is
\begin{equation}\label{finalintensity}
I = \int_\gamma \frac{f(r)^{3/2}} {   r^2} \sqrt{f(r)^{-1}+r^2 \left(\frac {d \psi}{dr}\right)^2}~ dr.
\end{equation}

\begin{figure}[h]
\centering 
\includegraphics[width=.5\textwidth]{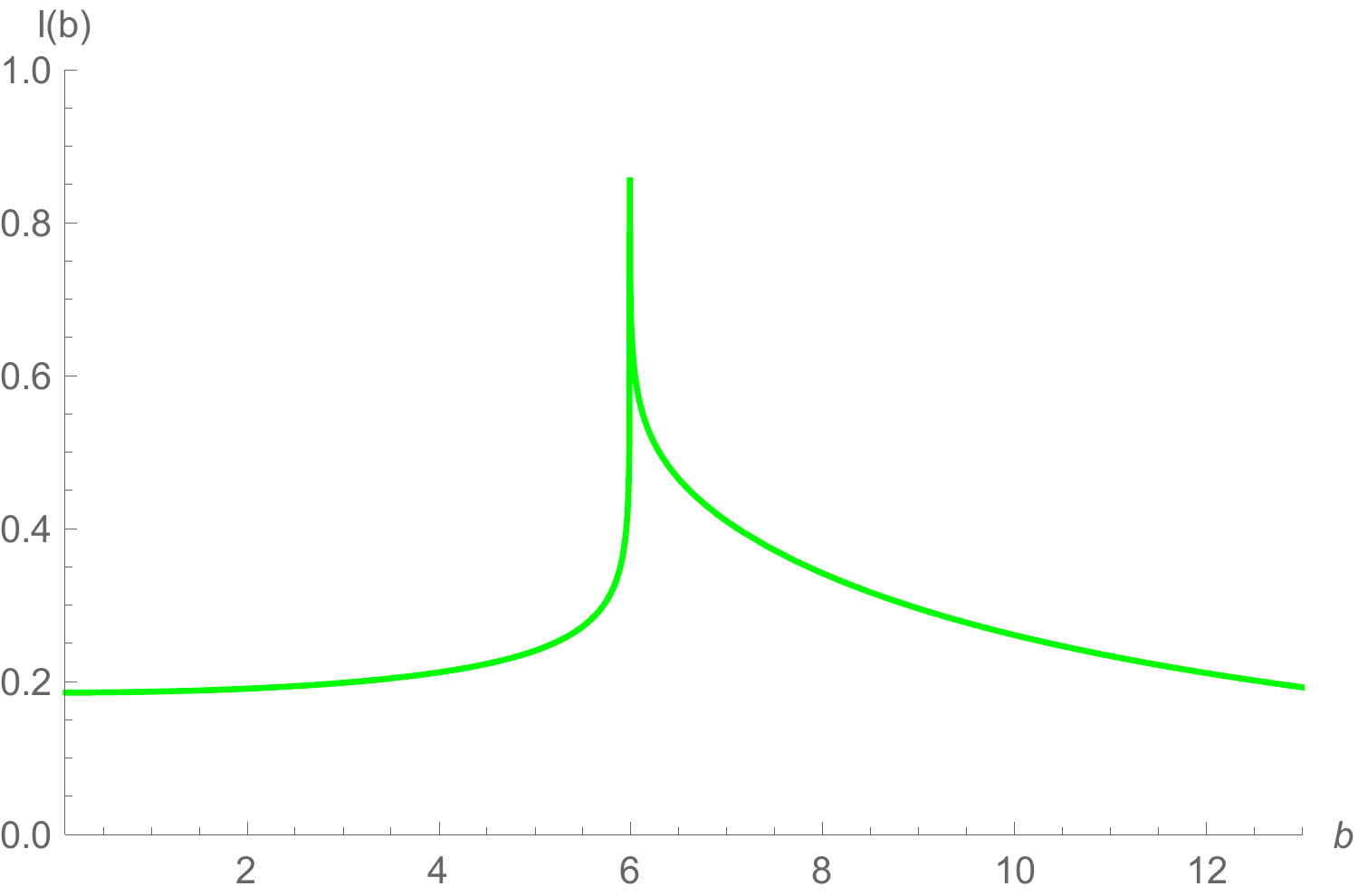}
\includegraphics[width=.4\textwidth]{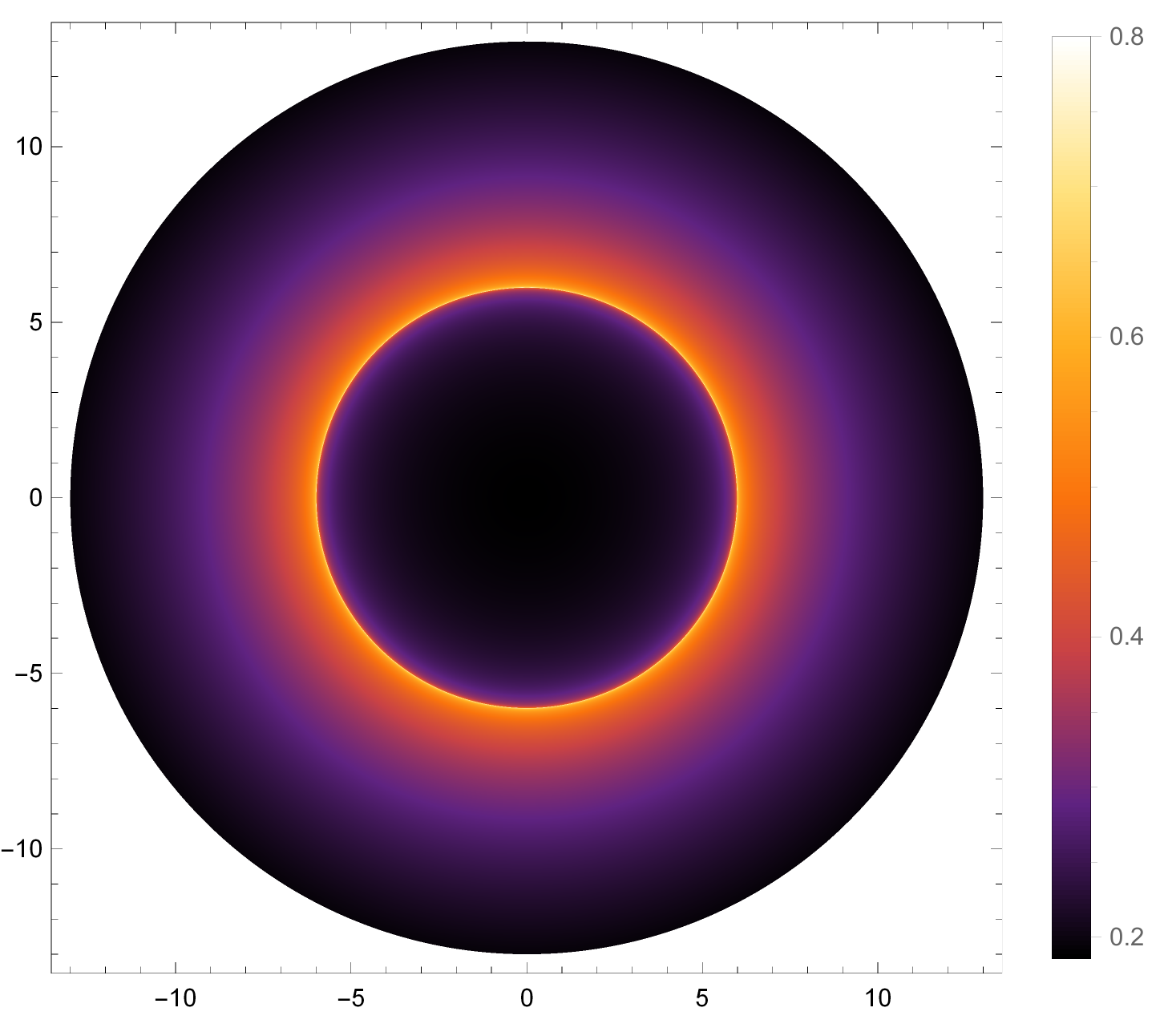}
\includegraphics[width=.5\textwidth]{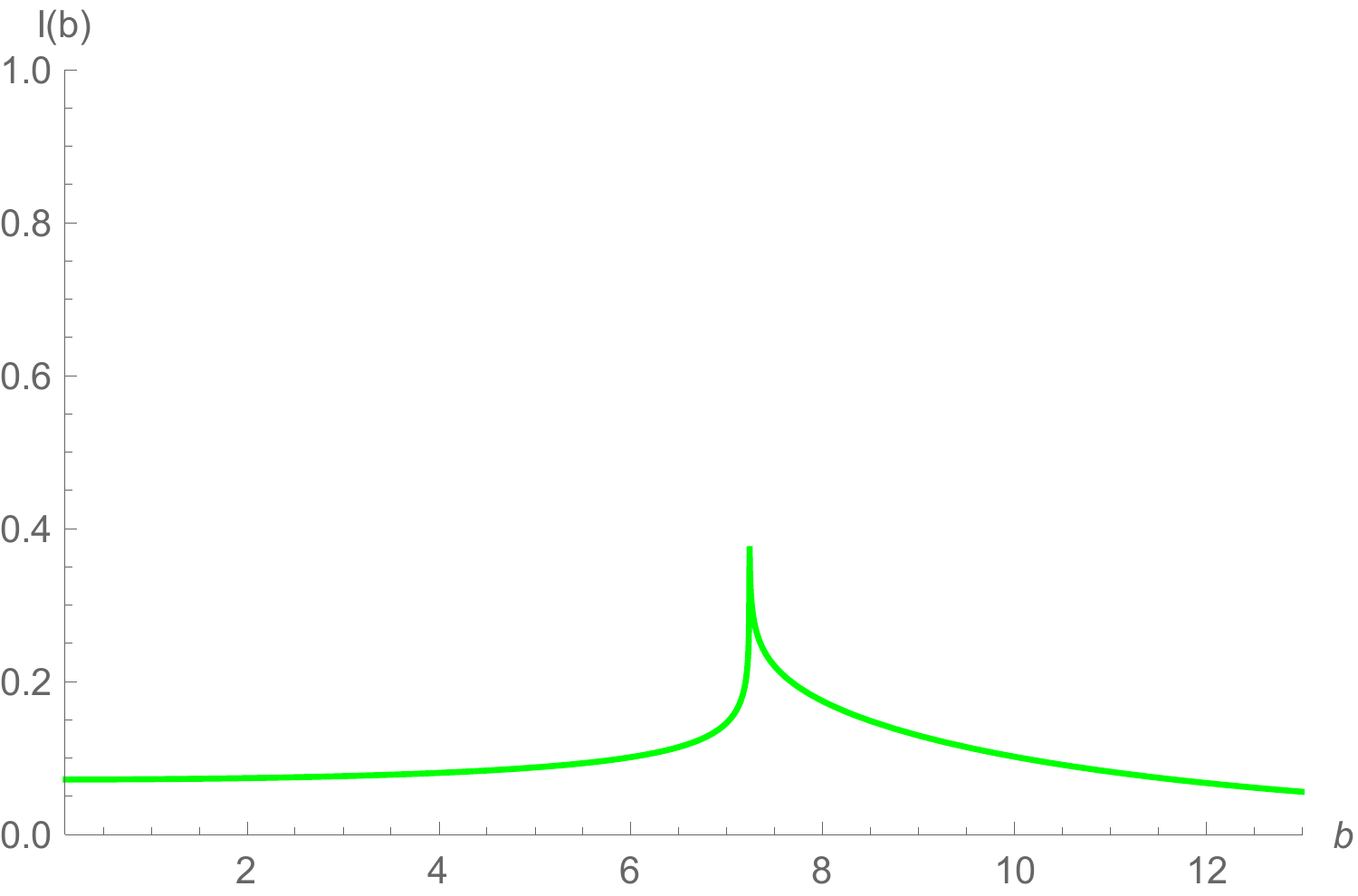}
\includegraphics[width=.4\textwidth]{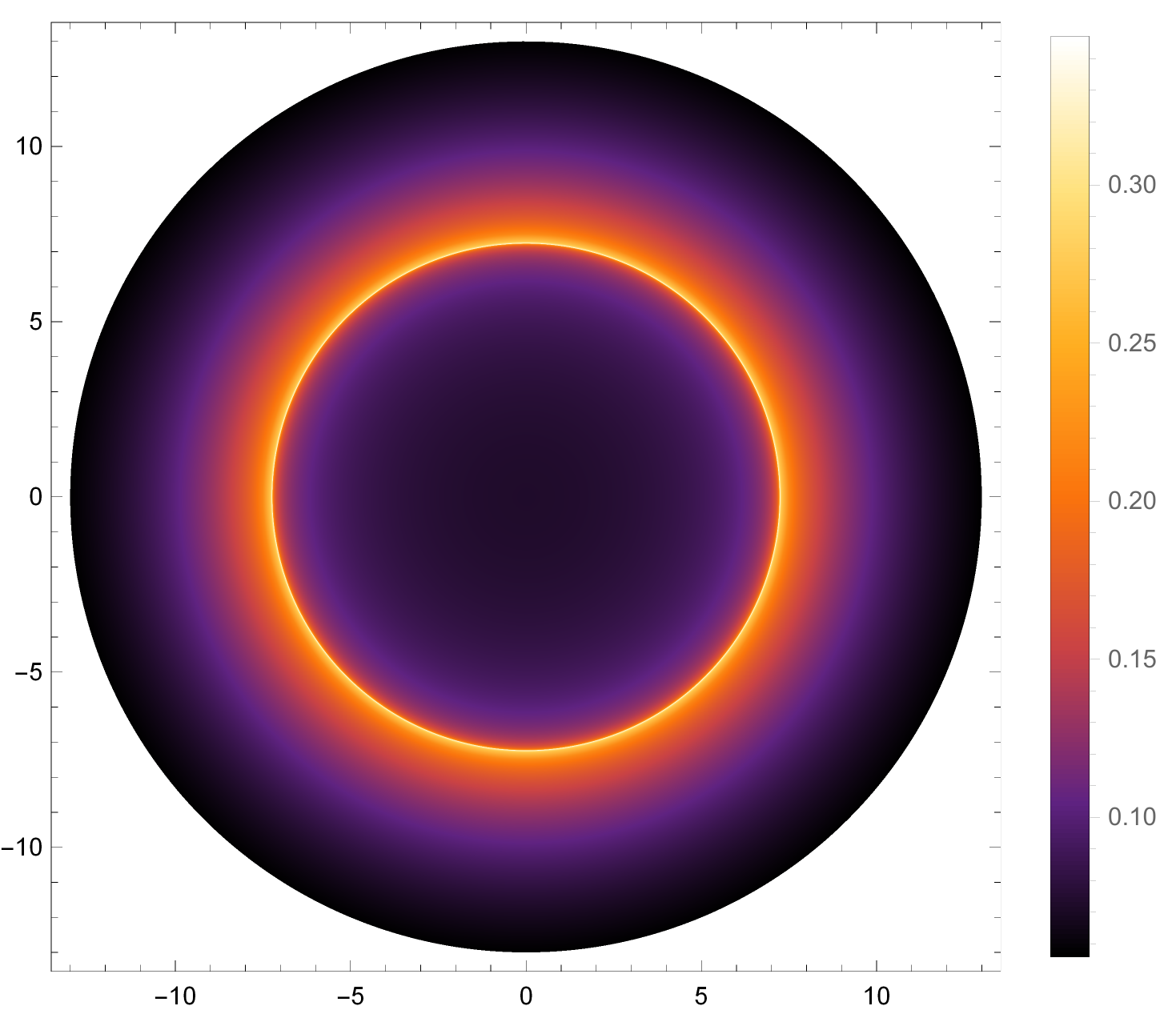}
\caption{\label{fig3} Profiles  of  the specific intensity $I(b)$ cast by a static spherical accretion,  viewed face-on by an observer near the cosmological horizon.  We set  $M=1$, $a=0.05$, $w=-0.5$ (top row) and  $w=-0.7$ (bottom row) as two examples. The details can be seen in the main texts.}
\end{figure}

Next we will employ Eq.(\ref{finalintensity}) to investigate the shadow of the quintessence black hole   with the static spherical accretion. Since the intensity depends on the
trajectory of the light ray, which is determined by the impact parameter $b$, we will investigate how the  intensity  varies with respect to the impact parameter  $b$. The observed specific intensities are shown in Figure \ref{fig3} with $w=-0.5$ (top row) and $w=-0.7$ (bottom row). From Figure \ref{fig3}, we see that as $b$ increases, the intensity ramps up first, then reaches a peak at $b_{ph}$ ($b_{ph}=5.98M$ for $w=-0.5$ and $b_{ph}=7.23M$ for $w=-0.7$), and finally drops rapidly to smaller values. This result is consistent with Figure \ref{fig1} and  Figure \ref{fig2}. Because for $b<b_{ph}$, the intensity originated from the accretion matter is absorbed  mostly by the black hole, so the observed intensity is little. For $b=b_{ph}$, the light ray revolves around the black hole several times,  thus the observed intensity is maximal. For $b>b_{ph}$, only the refracted light contributes to the  observed intensity. In particular, as $b$ becomes larger, the refracted light becomes less. The observed  intensity thus vanishes for large enough $b$.

From Figure \ref{fig3}, we can also see how the state parameter of dark energy will affect the observed intensity. For a fixed $a$ and $M$,  we find that the larger the state parameter is, the stronger the intensity will be, i.e., the intensity of $w=-0.5$ is stronger than that of $w=-0.7$. The 2-dimensional plot of the shadows are shown in the right column of Figure \ref{fig3}, with the impact parameter $b$ of the radius. We can see obviously that the shadow is circularly symmetric for the quintessence black hole, and outside the black hole, there is a bright ring, which is the photon sphere. The radius of the photon sphere for different $w$'s have been listed in Table \ref{tab11}. Obviously, the result in Figure \ref{fig3} is consistent with that in Table \ref{tab11}. That is, the larger the state parameter is, the smaller the  radius of the photon sphere will be.

In addition,  inside the photon sphere, we see that the specific intensity does not vanish
but has a small finite value. The reason is that part of the  radiation is escaped from the black hole. For $r<r_{ph}$, the solid angle of the
escaping radiation is  $2 \pi (1-\cos\theta)$, while for $r>r_{ph}$, the solid angle is  $2 \pi (1+\cos\theta)$, in which  $\theta$  is given by
\begin{equation}
\sin \theta =\frac{r_{ph}^{3/2}}{r} \left(1 - \frac{2 M} {r}- \frac{a}{ r^{ 3 w + 1} }\right)^{1/2}.
\end{equation}
In principle, one can calculate  the escaped net luminosity   by the following relation
\begin{equation}
L_{o}=\int_{r_h}^{r_{ph}} 4 \pi r^2 j (\nu_{\rm e}) 2 \pi (1-\cos\theta) dr+\int_{r_{ph}}^{\text{observer}} 4 \pi r^2 j (\nu_{\rm e}) 2 \pi (1+\cos\theta) dr.
\end{equation}
Obviously, for different  state parameters, the net luminosity observed by the static  observer in the quintessence black hole  is different, since solid angle is dependent on the state parameter.  This result is  consistent with the result in Figure \ref{fig3}.

\subsection{Shadows and photon spheres with an infalling  spherical accretion }

In this section, the optically thin accretion  is supposed to be infalling matters.  This infalling model is thought to be more realistic than the static accretion model, since most of  the accretion matters are dynamical in the universe. We still employ  Eq.(\ref{finalintensity}) to investigate the shadow cast by the infalling  accretion. Different from the static accretion, the redshift factor for the infalling accretion  is related to the velocity of the accretion, that is
\begin{equation}
g = \frac{k_\beta u^\beta_{\rm o}}{k_\gamma u^\gamma_{\rm e}} , \label{redf}
\end{equation}
in which $k^\mu=\dot{x_\mu}$ is the four-velocity of the photon, $u^\mu_{\rm o} = (1,0,0,0)$ is
the four-velocity of the  static observer, and $u^\mu_{\rm e}$ is the four-velocity of
the accretion under consideration, given by
\begin{eqnarray}
u^t_{\rm e} = \frac{1}{f(r)},~~
u^r_{\rm e} = - \sqrt{ 1 - f(r) },~~
u^\theta_{\rm e} = u^\psi_{\rm e} = 0.
\end{eqnarray}
The four-velocity  of the photon has been  obtained
previously from  Eq.(\ref{time}) to Eq.(\ref{radial}). From these equations, we know that  $k_t=1/b$ is a constant, and $k_r$ can be inferred
from the equation $k_\alpha k^\alpha = 0$, that is
\begin{equation}
\frac{k_r}{k_t} = \pm \frac{1}{f(r)}\sqrt{   1 - \frac{b^2 f(r)}{r^2} } , \label{krkt}
\end{equation}
in which  the  $+$/$-$ corresponds to the case that the photon gets close to/away from
the black hole. With Eq.(\ref{krkt}), the redshift factor in Eq.(\ref{redf}) can be simplified as
\begin{equation}
g =  \frac{1}{u^{t}_e+k_r/k_e u^r_e}, \label{sredf}
\end{equation}
which is different from the static accretion case.

\begin{figure}[tbp]
\centering 
\includegraphics[width=.5\textwidth]{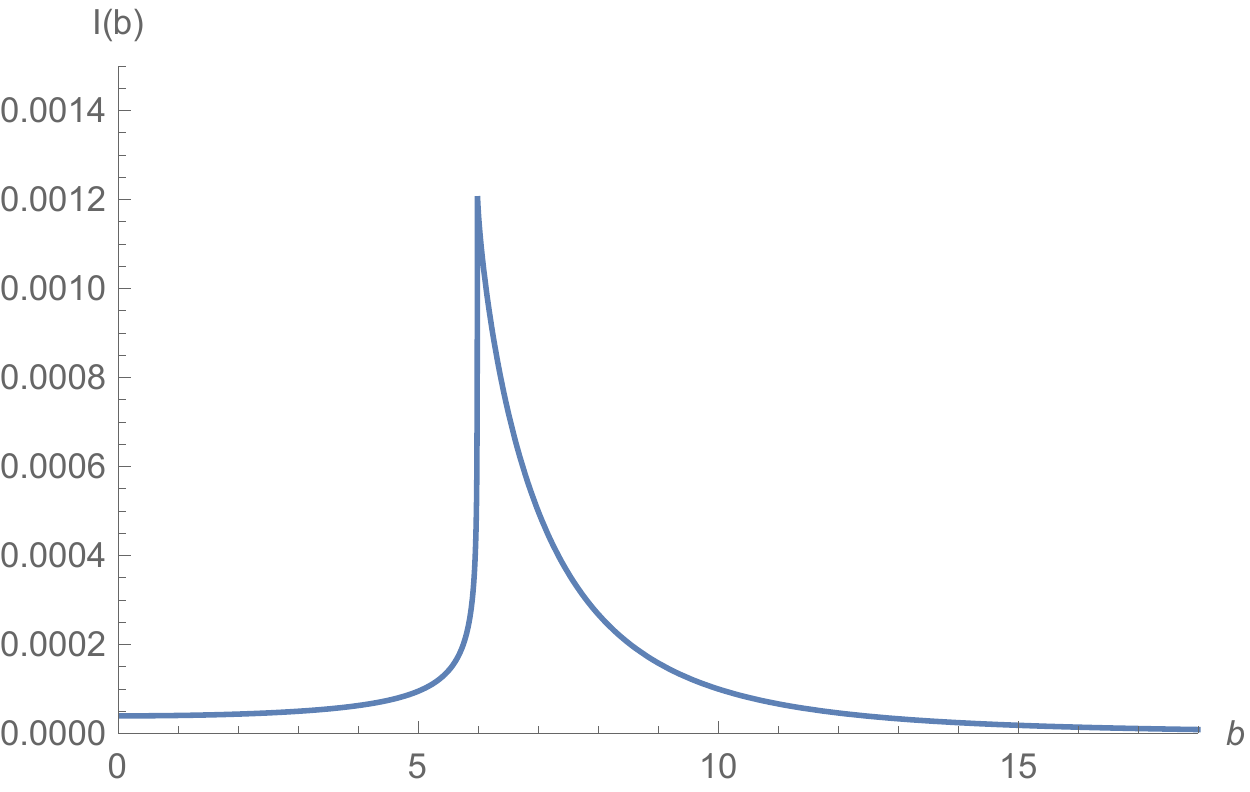}
\includegraphics[width=.4\textwidth]{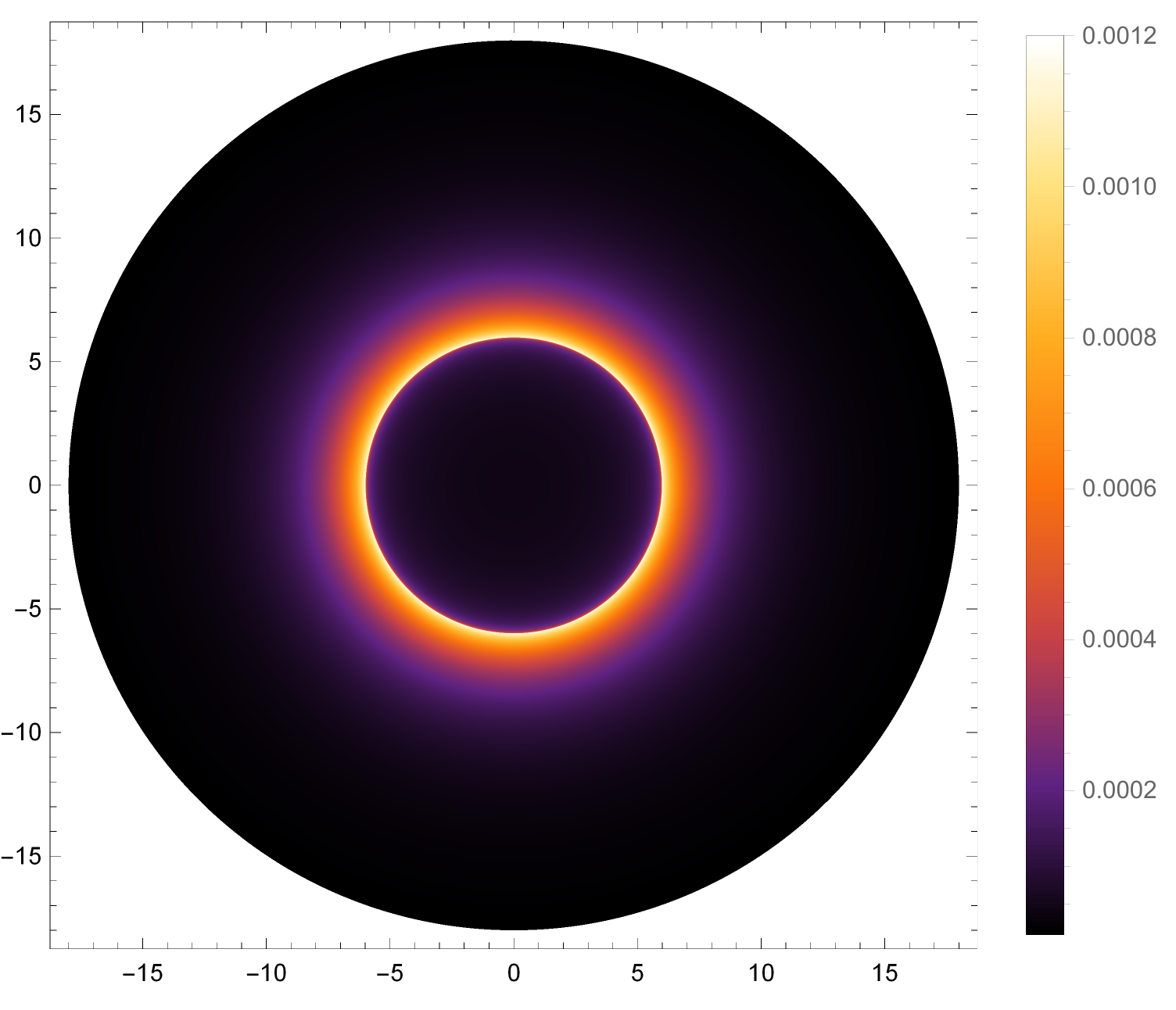}
\includegraphics[width=.5\textwidth]{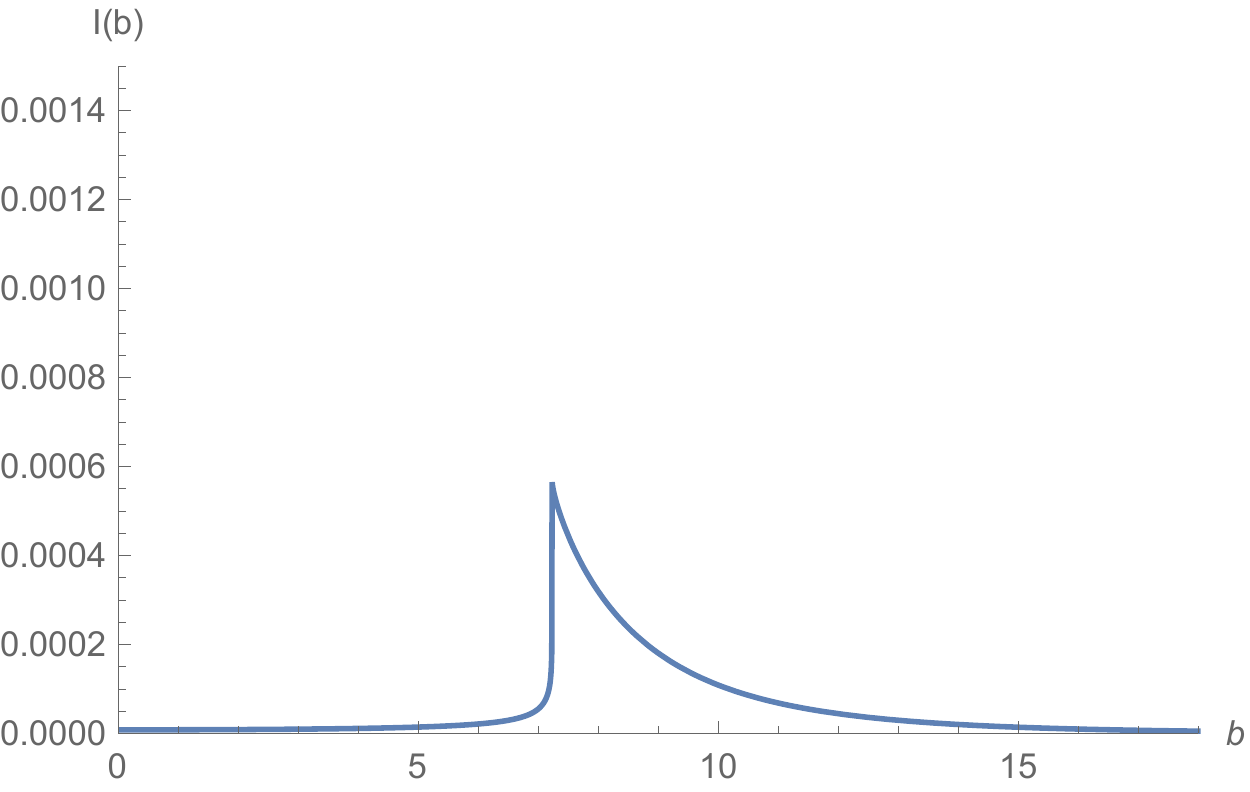}
\includegraphics[width=.4\textwidth]{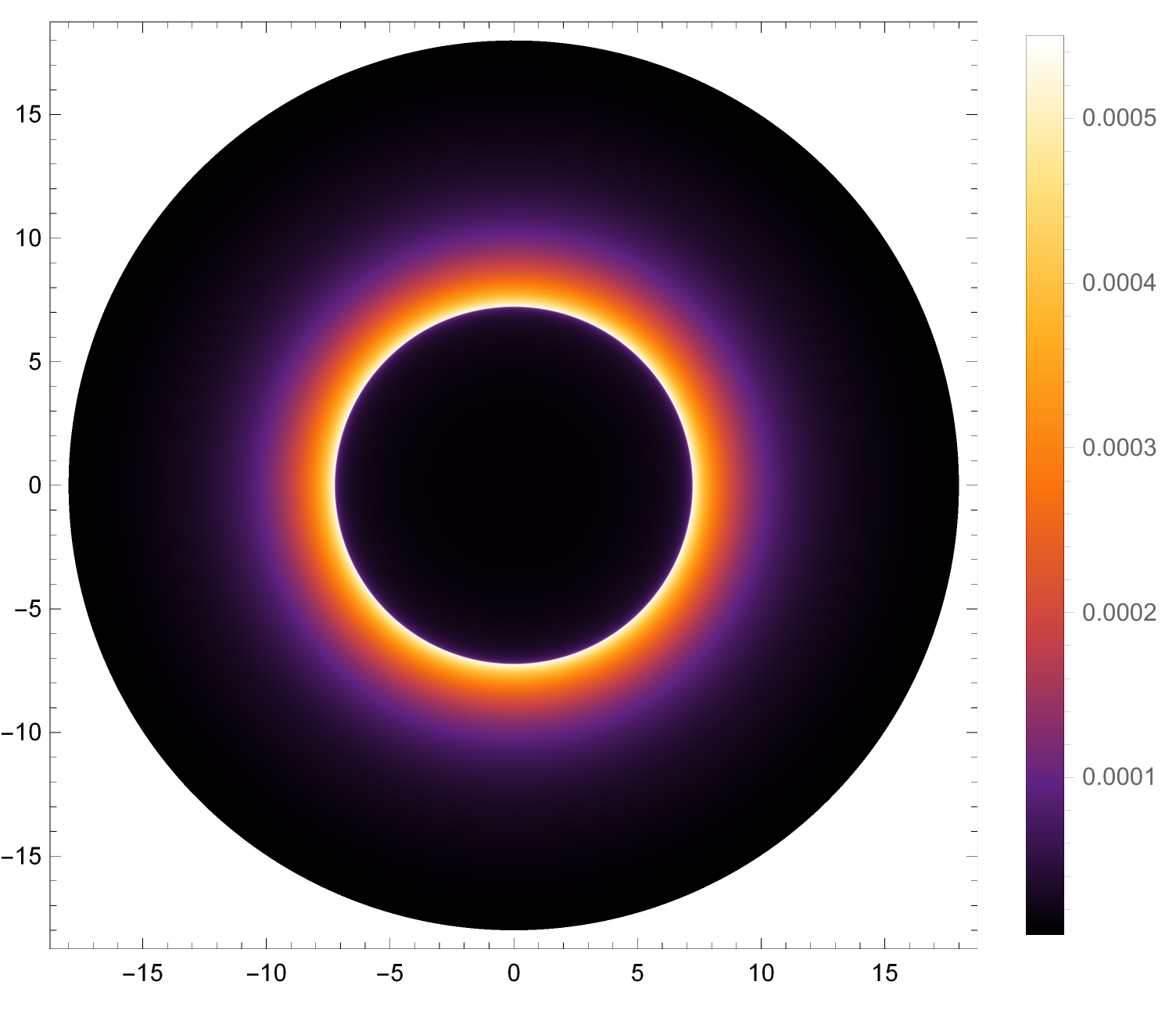}
\caption{\label{fig5} Profiles  of  the specific intensity $I(b)$ cast by an infalling spherical accretion,  viewed face-on by an observer near the cosmological horizon.  We set  $M=1$, $a=0.05$, $w=-0.5$ (top row) and  $w=-0.7$ (bottom row) as two examples. The details can be seen in the main texts.}
\end{figure}

In addition, the proper  distance  is defined by
\begin{equation}
dl_{\rm p} = k_\gamma u^\gamma_{\rm e}
ds=\frac{k_t}{g | k_r |} dr,
\end{equation}
where $s$ is the
affine parameter along the photon path $\gamma$.  With regard to the specific emissivity, we also
assume  that it is monochromatic, so that Eq.(\ref{profile}) is still valid. Integrating Eq.(\ref{intensity}) over the observed frequencies, we obtain
\begin{equation}
I \propto  \int_\gamma \frac{g^3 k_t dr}{r^2 | k_r |} .  \label{finaltensity}
\end{equation}
Now we will use  Eq.(\ref{finaltensity}) to investigate the shadow of the quintessence black hole numerically  with an infalling accretion.
Note that there is an  absolute sign for $ k_r$ in the denominator. Therefore, when the photon changes its motion direction, the sign before $ k_r$ should also change.  For different state parameter $w$'s, the observed intensity with respect to $b$ are  shown in Figure \ref{fig5}, in which the top row is for $w=-0.5$ while the bottom row is for $w=-0.7$. From  Figure \ref{fig5}, we find that as $b$ increases the intensity will increase as well to a peak $b=b_{ph}$, after the peak it drops to smaller values. This behavior is similar to that in the static accretion in Figure \ref{fig3}. We can also observe the effect of $w$ on the intensity, and find that the  intensity increases as  the value of  $ w$ increases, i.e., the intensity for $w=-0.5$ is stronger than $w=-0.7$.
The two dimensional image of the intensity is shown on the right column in Figure \ref{fig5}. We see that the radii of the shadows and the locations of the photon spheres are the same as the static case. That means the motion of the accretion does not affect the  radii of the shadows and the locations of the photon spheres. However, different from the static accretion,  the central region of the intensity for the infalling accretion is darker, which can be accounted for by the Doppler effect.  In particular, near the event horizon of the black hole, this effect is more obvious.

\section{Conclusions and discussions}
\label{conclusion}

We investigated the images of a black hole in the quintessence dark energy model. In particular, we studied the quintessence black hole surrounded by a geometrically and optically thin disk accretion, as well as a spherically symmetric accretion. We found that for the observed specific intensities, there existed a dark interior (shadow), lensing ring and photon ring for the thin disk accretion. But their positions depended on the profiles of the emission from the accretion near the black hole. Despite of this distinctions, the direct image of the accretion made a major contribution to the brightness of the black hole, while the lensing ring made a minor contribution and the photon ring made a negligible contribution. The effects of the quintessence state parameter was also important to the shadow images of the black hole, since this parameter would have impact on the distances between the observer and the event horizon.

We also studied the static and infalling spherically symmetric accretions, the images of the black hole would have a dark interior and a photon sphere. The infalling accretion would have a darker interior than the static case, which was due to the Doppler effect of the infalling matter. Different quintessence state parameter would change the positions of the photon spheres of the image. We expect that this would bring insights and implications for the quintessence dark energy model from the observations of the shadow images of the black holes in future EHT experiments.


\acknowledgments

This work is supported  by the National
Natural Science Foundation of China (Grant Nos. 11675140, 11705005, and 11875095).



\end{document}